\documentclass[a4paper,12pt]{article}
\usepackage{a4,amsmath,amssymb,axodraw,slashed,epsfig}
\usepackage[square,comma,numbers,sort&compress]{natbib}

\begin{document}


\newcommand{\bed}{\begin{displaymath}}
\newcommand{\eed}{\end{displaymath}}
\newcommand{\beq}{\begin{equation}}
\newcommand{\eeq}{\end{equation}}
\newcommand{\bea}{\begin{eqnarray}}
\newcommand{\eea}{\end{eqnarray}}
\newcommand{\tgb}{{\rm tg}\beta}
\newcommand{\tga}{{\rm tg}\alpha}
\newcommand{\stgb}{{\rm tg}^2\beta}
\newcommand{\sia}{\sin\alpha}
\newcommand{\coa}{\cos\alpha}
\newcommand{\sib}{\sin\beta}
\newcommand{\cob}{\cos\beta}
\newcommand{\MS}{\overline{\rm MS}}
\newcommand{\st}{\tilde{t}}
\newcommand{\sgl}{\tilde{g}}
\newcommand{\Dmb}{\ensuremath{\Delta_b}}

\newcommand{\tb}{{\rm tg}\beta}
\newcommand{\gluino}{\widetilde{g}}
\newcommand{\squark}{\tilde{q}}
\newcommand{\Mq}[1]{m_{\tilde{q}_{#1}}}
\newcommand{\mq}{m_{q}}
\newcommand{\Aq}{A_q}
\newcommand{\mix}{\widetilde{\theta}}
\newcommand{\Lag}{\mathcal{L}}
\newcommand{\lb}{\lambda_{b}}
\newcommand{\lt}{\lambda_t}
\newcommand{\Deltamb}{\Delta_b}
\newcommand{\DeltambQCD}{\Delta_b^{QCD}}
\newcommand{\DeltambELW}{\Delta_b^{elw}}
\newcommand{\mb}{m_{b}}
\newcommand{\mt}{m_{t}}
\newcommand{\bR}{b_R}
\newcommand{\bL}{b_L}
\newcommand{\At}{A_t}
\newcommand{\Sb}{\Sigma_b}
\newcommand{\Ao}[1]{A_0(#1)}
\newcommand{\Co}[5]{C_0(#1,#2;#3,#4,#5)}
\newcommand{\as}{\alpha_s}
\newcommand{\Mg}{m_{\tilde{g}}}
\newcommand{\Mb}[1]{m_{\tilde{b}_{#1}}}
\newcommand{\Mt}[1]{m_{\tilde{t}_{#1}}}
\newcommand{\NL}{\nonumber\\}
\newcommand{\bs}{\widetilde{b}}
\newcommand{\bsL}{\bs_L}
\newcommand{\bsR}{\bs_R}
\newcommand{\ts}{\widetilde{t}}
\newcommand{\tsR}{\widetilde{t}_R}
\newcommand{\tsL}{\widetilde{t}_L}
\newcommand{\higgsino}{\widetilde{h}}
\newcommand{\sbottom}{\tilde{b}}
\newcommand{\gh}{g_b^h}
\newcommand{\gH}{g_b^H}
\newcommand{\gA}{g_b^A}
\newcommand{\ght}{\tilde{g}_b^h}
\newcommand{\gHt}{\tilde{g}_b^H}
\newcommand{\gAt}{\tilde{g}_b^A}
\newcommand{\stb}{{\rm tg}^2\beta}
\newcommand{\Order}[1]{{\cal{O}}\left(#1\right)}
\newcommand{\Msusy}{M_{SUSY}}
\newcommand{\gluon}{g}
\newcommand{\sttop}{\widetilde{t}}
\newcommand{\CF}{C_F}
\newcommand{\CA}{C_A}
\newcommand{\TR}{T_R}
\newcommand{\Tr}[1]{{\rm Tr}\left[#1\right]}
\newcommand{\dk}{\frac{d^nk}{(2\pi)^n}}
\newcommand{\dqq}{\frac{d^nq}{(2\pi)^n}}
\newcommand{\partderiv}[1]{\frac{\partial}{\partial#1}}
\newcommand{\kmu}{k_{\mu}}
\newcommand{\qmu}{q_{\mu}}
\newcommand{\eUV}{\epsilon}
\newcommand{\dMq}[1]{\delta\Mq{#1}}
\newcommand{\gs}{g_s}
\newcommand{\MSbar}{\overline{\rm MS}}
\newcommand{\gPhit}{\tilde{g}_b^{\Phi}}
\newcommand{\muR}{\mu_{R}}
\newcommand{\bbbar}{b\bar{b}}
\newcommand{\MPhi}{M_{\Phi}}
\newcommand{\dlt}{\delta\lambda_t}
\newcommand{\GF}{{\rm G_F}}
\newcommand{\gPhi}{g_b^{\Phi}}
\newcommand{\mbMS}{\overline{m}_{b}}
\newcommand{\asrun}[1]{\as(#1)}
\newcommand{\NF}{N_F}
\newcommand{\gtPhi}{g_t^{\Phi}}
\newcommand{\Log}[1]{\log\left(#1\right)}
\newcommand{\smallaeff}{small $\alpha_{eff}$}
\newcommand{\ra}{\rightarrow}
\newcommand{\tautau}{\tau^+\tau^-}

\newcommand{\Quark}[4]{\ArrowLine(#1,#2)(#3,#4)}
\newcommand{\Gluino}[5]{\Gluon(#1,#2)(#3,#4){3}{#5}\Line(#1,#2)(#3,#4)}
\newcommand{\CrossedCircle}[6]{\BCirc(#1,#2){5}\Line(#3,#6)(#5,#4)\Line(#5,#6)(#3,#4)}
\newcommand{\Higgsino}[5]{\Photon(#1,#2)(#3,#4){3}{#5}\Line(#1,#2)(#3,#4)}
\newcommand{\myGluon}[5]{\Gluon(#1,#2)(#3,#4){3}{#5}}
\newcommand{\Squark}[4]{\DashArrowLine(#1,#2)(#3,#4){2}}
\newcommand{\Squarknoarrow}[4]{\DashLine(#1,#2)(#3,#4){2}}
\newcommand{\Quarknoarrow}[4]{\ArrowLine(#1,#2)(#3,#4)}


\vskip-1.0cm

\begin{flushright}
PSI--PR--10--01 \\
\end{flushright}

\begin{center}
{\large \sc Supersymmetric Higgs Yukawa Couplings to \\[0.3cm] Bottom Quarks at
next-to-next-to-leading Order} \\[1cm]
\end{center}

\begin{center}
{\sc David Noth$^{1,2}$ and Michael Spira$^1$} \\[0.8cm]

\begin{small}
{\it \small
$^1$ Paul Scherrer Institut, CH--5232 Villigen PSI, Switzerland \\
$^2$ Institut f\"ur Theoretische Physik, Z\"urich University, CH--8057
Z\"urich, Switzerland}
\end{small}
\end{center}


\begin{abstract}
\noindent
The effective bottom Yukawa couplings are analyzed for the minimal
supersymmetric extension of the Standard Model at two-loop accuracy
within SUSY--QCD. They include the resummation of the dominant
corrections for large values of $\rm tg\beta$. In particular the
two-loop SUSY-QCD corrections to the leading SUSY-QCD and top-induced
SUSY-electroweak contributions are addressed. The residual theoretical
uncertainties range at the per-cent level.  
\end{abstract}


\section{Introduction}
The Standard Model (SM) predicts the existence of one scalar Higgs boson
which constitutes the remainder of electroweak symmetry breaking by
means of the Higgs mechanism \cite{Higgs:1964ia}. In all experiments
this particle has escaped detection so far. Due to the hierarchy problem
in the context of Grand Unified Theories supersymmetric (SUSY)
extensions of the SM are considered as the most attractive solutions
\cite{susy}.  The minimal supersymmetric extension of the SM (MSSM)
requires the existence of five elementary Higgs bosons, two neutral
CP-even (scalar) bosons $h$, $H$, one neutral CP-odd (pseudoscalar)
boson $A$ and two charged bosons $H^\pm$. At lowest order all couplings
and masses of the MSSM Higgs sector are fixed by two independent input
parameters, which are generally chosen as $\tgb=v_2/v_1$, the ratio of
the two vacuum expectation values $v_{1,2}$, and the pseudoscalar Higgs
mass $M_A$.  Including the one-loop and dominant two-loop corrections
the upper bound on the light scalar Higgs mass is $M_h\lesssim 135$ GeV
\cite{mssmrad}.  More recent first three-loop results confirm this upper
bound within less than 1 GeV \cite{mssmrad3}.  The couplings of the
various Higgs bosons to fermions and gauge bosons depend on mixing
angles $\alpha$ and $\beta$, which are defined by diagonalizing the
neutral and charged Higgs mass matrices. They are collected in Table
\ref{tab:higgs_sm_couplings} relative to the SM Higgs couplings.
\begin{table}[t]
\begin{center}
\begin{tabular}{|lc||ccc|} \hline
\multicolumn{2}{|c||}{$\Phi$} & $g^\Phi_u$ & $g^\Phi_d$ &  $g^\Phi_V$ \\
\hline \hline
SM~ & $H$ & 1 & 1 & 1 \\ \hline
MSSM~ & $h$ & $\cos\alpha/\sin\beta$ & $-\sin\alpha/\cos\beta$ & $\sin(\beta-\alpha)$ \\
& $H$ & $\sin\alpha/\sin\beta$ & $\cos\alpha/\cos\beta$ & $\cos(\beta-\alpha)$ \\
& $A$ & $ 1/\tan\beta$ & $\tan\beta$ & 0 \\ \hline
\end{tabular}
\end{center}
\vspace*{-0.4cm}
\caption{MSSM Higgs couplings to SM particles relative to SM Higgs couplings.}
\label{tab:higgs_sm_couplings}
\end{table}
For large values of $\tgb$ the down-type Yukawa couplings are strongly
enhanced, while the up-type Yukawa couplings are suppressed, unless the
light (heavy) scalar Higgs mass ranges at its upper (lower) bound, where
the couplings become SM-like. This feature causes the dominance of
bottom-Yukawa-coupling induced processes for large values of $\tgb$ at
present and future colliders as Higgs decays into bottom quarks and
Higgs bremsstrahlung off bottom quarks at hadron and $e^+e^-$ colliders.
Moreover, Higgs boson production via gluon fusion $gg\to h,H,A$ is
dominated by the bottom-loop contributions for large $\tgb$. Thus, the
bottom Yukawa coupling determines  the profile of the MSSM Higgs bosons
for large $\tgb$ to a large extent.

The negative direct searches for the MSSM Higgs bosons at LEP2 yield
lower bounds of $M_{h,H} > 92.8$ GeV and $M_A > 93.4$ GeV. The range
$0.7 < \tgb < 2.0$ in the MSSM is excluded by the Higgs searches for a
SUSY scale $M_{SUSY}=1$ TeV at the LEP2 experiments \cite{lep2}.
Presently and in the future Higgs bosons can be searched for at the
Tevatron at Fermilab \cite{Carena:2000yx}, a proton-antiproton collider
with a center-of-mass energy of 1.96 TeV, and the proton-proton collider
LHC (Large Hadron Collider) with 14 TeV center-of-mass energy
\cite{atlas_cms_tdrs} as well as a future linear $e^+e^-$ collider with
a center-of-mass energy up to about 1 TeV \cite{ilc}.

The mass degeneracy of the fermions $f$ and their superpartners
$\tilde{f}$ is removed by the soft SUSY breaking terms, which induce
mixing of the current eigenstates $\tilde{f}_L$ and $\tilde{f}_R$ in
addition.  The sfermion mass matrix in the current eigenstate basis is
given by\footnote{For simplicity, the $D$-terms have been absorbed in
the sfermion mass parameters $M_{\tilde f_{L/R}}^2$.}
\beq
{\cal M}^2_{\tilde f}  = \left(\begin{array}{cc}M_{LL}^2 &
M_{LR}^2 \\M_{RL}^2 & M_{RR}^2 \\\end{array} \right) =
\left(\begin{array}{cc} M_{\tilde f_L}^2 + m_f^2\ & m_f(A_f-\mu r_f)
\\ m_f(A_f-\mu r_f)\ & M_{\tilde f_R}^2 + m_f^2 \\\end{array}\right)
\label{eq:MLL_MRR_MRL}
\eeq
with the parameters $r_d=1/r_u=\tb$ for down- and up-type sfermions.
The mass eigenstates $\tilde f_{1,2}$ of the sfermions $\tilde f$ are
related to the current eigenstates $\tilde f_{L,R}$ by mixing angles
$\theta_f$,
\begin{eqnarray}
\tilde f_1 & = & \tilde f_L \cos\theta_f + \tilde f_R \sin \theta_f
\nonumber \\
\tilde f_2 & = & -\tilde f_L\sin\theta_f + \tilde f_R \cos \theta_f \, ,
\label{eq:sfmix}
\end{eqnarray}
which are proportional to the masses of the ordinary fermions. Thus
mixing effects are only important for the third-generation sfermions
$\tilde t, \tilde b, \tilde \tau$.  The mixing angles acquire the form
\begin{equation}
\sin 2\theta_f = \frac{2m_f (A_f-\mu r_f)}{m_{\tilde f_1}^2 - m_{\tilde
f_2}^2}
~~~,~~~
\cos 2\theta_f = \frac{M_{\tilde f_L}^2 - M_{\tilde f_R}^2}{m_{\tilde
f_1}^2
- m_{\tilde f_2}^2}
\end{equation} 
and the masses of the squark mass eigenstates are given by
\begin{equation}
m_{\tilde f_{1,2}}^2 = m_f^2 + \frac{1}{2}\left[ M_{\tilde f_L}^2 +
M_{\tilde f_R}^2 \mp \sqrt{(M_{\tilde f_L}^2 - M_{\tilde f_R}^2)^2 +
4m_f^2 (A_f - \mu r_f)^2} \right] \, .
\end{equation}

The topic of this paper is the calculation of the NNLO SUSY-QCD and
top-induced SUSY-electroweak corrections to the effective bottom Yukawa
couplings. These results will affect all processes to which the bottom
Yukawa couplings contribute, i.e.~in particular the Higgs decay widths
and Higgs radiation off bottom quarks at hadron colliders which
constitutes the dominant Higgs boson production channel for large $\tb$
at the Tevatron and the LHC \cite{spira:98}.

\section{Effective Bottom Yukawa Couplings}
The leading parts of the SUSY--QCD and SUSY--electroweak corrections to
bottom Yukawa coupling induced processes can be absorbed in effective
bottom Yukawa couplings. These contributions correspond to the limit of
heavy supersymmetric particle masses compared to the typical energy
scales of the particular process. The accuracy of this heavy mass
approximation has been investigated for neutral MSSM Higgs decays into
bottom quarks $h/H/A\to b\bar b$ \cite{GHS}, charged Higgs decays to
top and bottom quarks $H^\pm\to tb$ \cite{CGNW} and Higgs radiation off
bottom quarks at $e^+e^-$ colliders \cite{ee2hbb} and hadron colliders
\cite{pp2chtb, pp2hbb} by comparing it to the full NLO results. For
large values of $\tgb$ the approximation turns out to agree with the NLO
results to better than one per cent.

\subsection{Effective Lagrangian}
The leading corrections can be obtained from the effective Lagrangian
\cite{GHS,CGNW}
\bea
{\cal L}_{eff} & = & -\lambda_b \overline{b_R} \left[ \phi_1^0
+ \frac{\Delta_b}{\tgb} \phi_2^{0*} \right] b_L + h.c. \nonumber \\
& = & -m_b \bar b \left[1+i\gamma_5 \frac{G^0}{v}\right] b
-\frac{m_b/v}{1+\Delta_b} \bar b \left[ g_b^h \left(
1-\frac{\Delta_b}{\tga~\tgb}\right) h \right. \nonumber \\
& & \hspace*{2cm} \left. + g_b^H \left( 1+\Delta_b
\frac{\tga}{\tgb}\right) H
- g_b^A \left(1-\frac{\Delta_b}{\stgb} \right) i \gamma_5 A \right] b
\label{eq:leff}
\eea
with the one-loop expressions ($C_F= 4/3$) \cite{Hall:1993gn}
\bea
\Delta_b^{(1)} & = & \Delta_b^{QCD\,(1)} + \Delta_b^{elw\,(1)}
\nonumber \\
\Delta_b^{QCD\,(1)} & = &
\frac{C_F}{2}~\frac{\alpha_s(\mu_R)}{\pi}~m_{\sgl}~\mu~\tgb~
I(m^2_{\sbottom_1},m^2_{\sbottom_2},m^2_{\sgl}) \nonumber \\
\Delta_b^{elw\,(1)} & = & \frac{\lambda_t^2(\mu_R)}{(4\pi)^2}~A_t~\mu~\tgb~
I(m_{\st_1}^2,m_{\st_2}^2,\mu^2)
\label{eq:effpar}
\eea
The generic function $I$ is defined as
\beq
I(a,b,c) = \frac{\displaystyle ab\log\frac{a}{b} + bc\log\frac{b}{c}
+ ca\log\frac{c}{a}}{(a-b)(b-c)(a-c)}
\eeq
The fields $\phi_1^0$ and $\phi_2^0$ denote the neutral components of
the Higgs doublets coupling to down- and up-type quarks, respectively.
They are related to the mass eigenstates $h,H,A$ by
\bea
\phi_1^0 & = & \frac{1}{\sqrt{2}}\left[ v_1 + H\coa - h\sia + iA\sib -
iG^0\cob
\right] \nonumber \\
\phi_2^0 & = & \frac{1}{\sqrt{2}}\left[ v_2 + H\sia + h\coa + iA\cob +
iG^0\sib
\right]
\eea
The two vacuum expectation values are related to the Fermi constant
$G_F$, $v^2={v_1^2+v_2^2} = 1/{\sqrt{2}G_F}$. The would-be Goldstone
field $G^0$ is absorbed by the $Z$ boson and generates its longitudinal
component. The top Yukawa coupling $\lambda_t$ is related to the top
mass by $m_t = \lambda_t v_2/\sqrt{2}$ at lowest order. The corrections
$\Dmb$ induce a modification of the relation between the bottom quark
mass $m_b$ and the bottom Yukawa coupling $\lambda_b$,
\beq
m_b = \frac{\lambda_b v_1}{\sqrt{2}} \left[ 1 + \Delta_b \right]
\eeq
The effective Lagrangian of Eq.~(\ref{eq:leff}) can be parametrized as
\beq
{\cal L}_{eff} = -\frac{m_b}{v}\ \bar{b}\ \big[\ \tilde g^h_b\ h +
\tilde g^H_b\ H - \tilde g^A_b\ i\gamma_5\ A\ ]\ b
\eeq
with the effective (resummed) couplings
\bea
\tilde g^h_b & = & \frac{g^h_b}{1+\Delta_b}\left[ 1 -
\frac{\Delta_b}{\tga\tgb}  \right] \nonumber \\
\tilde g^H_b & = & \frac{g^H_b}{1+\Delta_b}\left[ 1 + \Delta_b
\frac{\tga}{\tgb} \right] \nonumber \\
\tilde g^A_b & = & \frac{g^A_b}{1+\Delta_b}\left[ 1 -
\frac{\Delta_b}{\tgb^2} \right]
\label{eq:rescoup}
\eea
Although the SUSY corrections $\Delta_b$ are loop suppressed, they turn
out to be significant for large values of $\tgb$ and moderate or large
$\mu$ values. In these cases they constitute the dominant supersymmetric
radiative corrections to the bottom Yukawa couplings.  It should be
noted that the effective Lagrangian in Eq.~(\ref{eq:leff}) has been
derived by integrating out the heavy SUSY particles so that it is not
restricted to large values of $\tgb$ only. In order to improve the
perturbative result it has been shown that the Lagrangian of
Eq.(\ref{eq:leff}) resums all terms of ${\cal
O}\left[(\alpha_s\,\mu\,\tgb)^n\right]$ and ${\cal
O}\left[(\lambda^2_t\,A_t\,\tgb)^n\right]$ \cite{CGNW}. The additional
resummation of terms proportional to $\alpha_s A_b$ \cite{GHS} will be
neglected in this paper.

\subsection{Low Energy Theorems}
The determination of higher-order corrections to the effective bottom
Yukawa couplings would usually require the calculation of the
corresponding three-point functions. This calculation can be reduced to
the evaluation of self-energy diagrams by the use of low energy theorems
\cite{let}. The basic idea is that any matrix element with an external
Higgs boson can be related to the analogous matrix element without the
external Higgs particle in the limit of vanishing Higgs momentum by the
simple replacements $v_1\to \sqrt{2}\phi_1^0$ and $v_2\to
\sqrt{2}\phi_2^{0*}$ in the latter. Thus we only need to compute the
corresponding pieces of the bottom quark self-energy. The leading pieces
$\Delta_b$ emerge from the scalar part $\Sigma_S(m_b)$ of the
self-energy\footnote{The fermionic self-energy can be decomposed into a
scalar, vectorial and axial-vectorial part according to $\Sigma(p) =
\Sigma_S(p) + \slashed{p}\,\Sigma_V(p) +
\slashed{p}\gamma_5\,\Sigma_A(p)$.} giving rise to the following
relation between the pole mass $m_b$ of the bottom quark and the bottom
Yukawa coupling $\lambda_b$\footnote{The bottom Yukawa coupling is not
renormalized at ${\cal O}(\alpha_s\mu\tgb)$ and ${\cal
O}(\lambda_t^2\,A_t\,\tgb)$ in the effective Lagrangian, but receives
only non-leading contributions in $\tgb$ \cite{GHS}, which do not
contribute to the effective Lagrangian.},
\beq
m_b=\frac{\lambda_b}{\sqrt{2}}v_1 + \Sigma_S(m_b)
\label{eq:sigma_s}
\eeq
where the leading terms of the self-energy $\Sigma_S(m_b)$ for large
values of $\tgb$ are given by
\bea
\Sigma_S(m_b) & = & \frac{\lambda_b}{\sqrt{2}}\,v_1\ \Delta_b
\nonumber \\
\Delta_b & = & \Delta_b^{QCD} + \Delta_b^{elw}
\eea
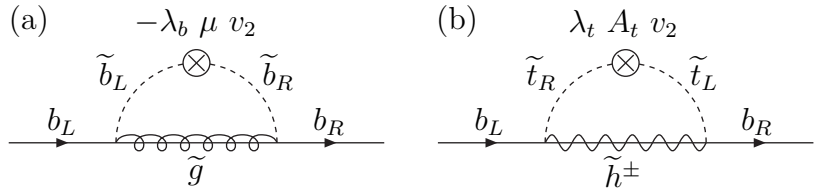
\begin{figure}[t]
\begin{center}
\begin{picture}(300,60)(0,0)
\Text(0,40)[lb]{(a)}
\Quark{0}{0}{40}{0}
\Gluino{40}{0}{100}{0}{6}
\Quark{100}{0}{140}{0}
\DashCArc(70,0)(30,0,180){2}
\CrossedCircle{70}{30}{67}{27}{73}{33}
\Text(20,3)[cb]{$\bL$}
\Text(120,3)[cb]{$\bR$}
\Text(45,20)[rb]{$\bsL$}
\Text(95,20)[lb]{$\bsR$}
\Text(70,-5)[ct]{$\gluino$}
\Text(70,40)[cb]{$-\lb\ \mu\ v_2$}
\Text(160,40)[lb]{(b)}
\Quark{160}{0}{200}{0}
\Higgsino{200}{0}{260}{0}{6}
\Quark{260}{0}{300}{0}
\DashCArc(230,0)(30,0,180){2}
\CrossedCircle{230}{30}{227}{27}{233}{33}
\Text(180,3)[cb]{$\bL$}
\Text(280,3)[cb]{$\bR$}
\Text(205,20)[rb]{$\tsR$}
\Text(255,20)[lb]{$\tsL$}
\Text(230,-5)[ct]{$\higgsino^\pm$}
\Text(230,40)[cb]{$\lt\ \At\ v_2$}
\end{picture}
\end{center}
\caption{\it One-loop diagrams of (a) the SUSY-QCD and (b) the
top-induced SUSY-electroweak contributions to the bottom self-energy
with the off-diagonal mass insertions corresponding to the corrections
$\Delta_b$ of the bottom Yukawa couplings. The virtual particles involve
bottom quarks $b$ and squarks $\sbottom$, top squarks $\st$, gluinos
$\gluino$ and charged Higgsinos $\higgsino^\pm$.}
\label{fig:1loop_Selfenergies}
\end{figure}
The terms $\Delta_b^{QCD}$ and $\Delta_b^{elw}$ can be derived from
off-diagonal mass insertions of the type $-\lambda_b\mu v_2$ in the
virtual sbottom propagators and of the type $\lambda_t A_t v_2$ in the
virtual stop propagators, as is illustrated in Figure
\ref{fig:1loop_Selfenergies} at one-loop level. The result of these
diagrams is given by the finite expressions of Eq.(\ref{eq:effpar})
after rotation of the fields in the current eigenstate basis to the mass
eigenstates. These expressions are not renormalized since there is no
tree-level coupling of bottom quarks to the Higgs field $\phi_2^{0*}$.
By means of power-counting arguments it can be proven that
Eq.~(\ref{eq:leff}) includes and resums all leading terms of ${\cal
O}\left[(\alpha_s\,\mu\,\tgb)^n\right]$ and ${\cal
O}\left[(\lambda_t^2\,A_t\,\tgb)^n\right]$ \cite{CGNW,GHS}.

\section{NNLO Corrections}
The calculation of the NNLO corrections to the effective bottom Yukawa
couplings requires the determination of the leading NNLO corrections to
the bottom self-energy. The full NNLO results of the bottom quark
self-energy have been presented in Refs.~\cite{bednyakov, martin,
mihaila} before. Our results for the self-energy at NNLO can be
extracted from their calculation by keeping only the terms leading in
$\tgb$ in principle. The renormalization, however, has to be adjusted to
our scheme in order to compare the final results. The results of
Ref.~\cite{bednyakov} are only shown for equal SUSY-breaking squark
masses $M^2_{\tilde q_L}=M^2_{\tilde q_R}$ for the full relation between
the pole quark mass and the $\overline{DR}$ quark mass.  However, since
the self-energy is a symmetric function of both sbottom masses
$m_{\tilde b_1}$ and $m_{\tilde b_2}$ their results for the {\it linear}
$a_b$ terms can be compared to our results in the equal SUSY mass case.
Moreover, our calculation of the bottom self-energy has been performed
for the leading term for small bottom momenta so that we can compare
directly with the results of Ref.~\cite{mihaila} which displays results
in the same limit.

\subsection{Two-Loop Diagrams}
\begin{figure}[htbp]
\SetScale{0.8}
\begin{picture}(200,40)(-20,0)
\ArrowLine(0,0)(50,0) 
\Line(50,0)(150,0) 
\Gluon(50,0)(100,0){5}{3}
\Gluon(100,0)(150,0){5}{3}
\Gluon(100,50)(100,0){5}{3}
\ArrowLine(150,0)(200,0) 
\DashCArc(100,0)(50,0,180){5}
\put(10,6){$b$}
\put(148,6){$b$}
\put(88,16){$g$}
\put(56,-15){$\sgl$}
\put(100,-15){$\sgl$}
\put(38,25){$\sbottom$}
\put(116,25){$\sbottom$}
\put(0,50){$(a)$}
\end{picture}
\begin{picture}(200,40)(-10,0)
\ArrowLine(0,0)(50,0) 
\Line(50,0)(100,0) 
\Gluon(50,0)(100,0){5}{3}
\ArrowLine(100,0)(150,0)
\DashLine(100,50)(100,0){5}
\ArrowLine(150,0)(200,0) 
\GlueArc(100,0)(50,0,90){5}{5}
\DashCArc(100,0)(50,90,180){5}
\put(10,6){$b$}
\put(148,6){$b$}
\put(86,16){$\sbottom$}
\put(56,-15){$\sgl$}
\put(100,-12){$b$}
\put(44,32){$\sbottom$}
\put(116,32){$g$}
\end{picture} \\
\begin{picture}(200,90)(-20,0)
\ArrowLine(0,0)(50,0) 
\ArrowLine(50,0)(100,0) 
\Line(100,0)(150,0)
\Gluon(100,0)(150,0){5}{3}
\DashLine(100,50)(100,0){5}
\ArrowLine(150,0)(200,0) 
\GlueArc(100,0)(50,90,180){5}{5}
\DashCArc(100,0)(50,0,90){5}
\put(10,6){$b$}
\put(148,6){$b$}
\put(86,16){$\sbottom$}
\put(56,-12){$b$}
\put(100,-15){$\sgl$}
\put(40,32){$g$}
\put(116,32){$\sbottom$}
\end{picture}
\begin{picture}(200,90)(-10,0)
\ArrowLine(0,0)(50,0) 
\Line(50,0)(100,0) 
\Line(100,50)(100,0) 
\Gluon(50,0)(100,0){5}{3}
\Gluon(100,0)(150,0){5}{3}
\Gluon(100,50)(100,0){5}{3}
\ArrowLine(150,0)(200,0) 
\DashCArc(100,0)(50,90,180){5}
\ArrowArcn(100,0)(50,90,0)
\put(10,6){$b$}
\put(148,6){$b$}
\put(88,16){$\sgl$}
\put(56,-15){$\sgl$}
\put(100,-15){$g$}
\put(38,25){$\sbottom$}
\put(116,25){$b$}
\end{picture} \\
\begin{picture}(200,90)(-10,0)
\ArrowLine(0,0)(50,0) 
\Line(100,0)(150,0) 
\Line(100,50)(100,0) 
\Gluon(50,0)(100,0){5}{3}
\Gluon(100,0)(150,0){5}{3}
\Gluon(100,50)(100,0){5}{3}
\ArrowLine(150,0)(200,0) 
\DashCArc(100,0)(50,0,90){5}
\ArrowArcn(100,0)(50,180,90)
\put(10,6){$b$}
\put(148,6){$b$}
\put(88,16){$\sgl$}
\put(56,-15){$g$}
\put(100,-15){$\sgl$}
\put(38,25){$b$}
\put(116,25){$\sbottom$}
\end{picture}
\begin{picture}(200,90)(-20,0)
\ArrowLine(0,0)(50,0) 
\Line(50,0)(100,0) 
\Gluon(50,0)(100,0){5}{3}
\DashLine(100,0)(150,0){5}
\ArrowLine(100,50)(100,0)
\ArrowLine(150,0)(200,0) 
\CArc(100,0)(50,0,90)
\GlueArc(100,0)(50,0,90){5}{5}
\DashCArc(100,0)(50,90,180){5}
\put(10,6){$b$}
\put(148,6){$b$}
\put(86,16){$b$}
\put(56,-15){$\sgl$}
\put(100,-15){$\sbottom$}
\put(44,32){$\sbottom$}
\put(116,32){$\sgl$}
\end{picture} \\
\begin{picture}(200,90)(-10,0)
\ArrowLine(0,0)(50,0) 
\ArrowLine(150,0)(200,0) 
\ArrowLine(50,0)(70,0) 
\ArrowLine(130,0)(150,0) 
\Line(70,0)(130,0) 
\Gluon(70,0)(130,0){5}{4}
\GlueArc(100,0)(50,0,180){5}{10}
\DashCArc(100,0)(30,0,180){5}
\put(10,6){$b$}
\put(148,6){$b$}
\put(48,-12){$b$}
\put(76,-16){$\sgl$}
\put(112,-12){$b$}
\put(76,12){$\sbottom$}
\put(116,34){$g$}
\end{picture}
\begin{picture}(200,90)(-10,0)
\ArrowLine(0,0)(50,0)
\Line(50,0)(150,0)
\Gluon(50,0)(150,0){5}{7}
\ArrowLine(150,0)(200,0)
\DashCArc(100,0)(50,0,180){5}
\GOval(100,46)(10,20)(0){0.5}
\put(10,6){$b$}
\put(148,6){$b$}
\put(72,-15){$\sgl$}
\put(36,25){$\sbottom$}
\put(116,25){$\sbottom$}
\end{picture} \\
\begin{picture}(200,90)(-10,0)
\ArrowLine(0,0)(50,0)
\Line(50,0)(150,0)
\Gluon(50,0)(150,0){5}{7}
\ArrowLine(150,0)(200,0)
\DashCArc(100,0)(50,0,180){5}
\GlueArc(100,3.5)(25,0,180){5}{5}
\put(10,6){$b$}
\put(148,6){$b$}
\put(47,-14){$\sgl$}
\put(75,-14){$\sgl$}
\put(107,-14){$\sgl$}
\put(76,30){$g$}
\put(44,32){$\sbottom$}
\end{picture}
\begin{picture}(200,90)(-10,0)
\ArrowLine(0,0)(50,0)
\Line(50,0)(150,0)
\Gluon(50,0)(80,0){5}{2}
\ArrowLine(80,0)(120,0)
\DashCArc(100,0)(20,0,180){5}
\Gluon(120,0)(150,0){5}{2}
\ArrowLine(150,0)(200,0)
\DashCArc(100,0)(50,0,180){5}
\put(10,6){$b$}
\put(148,6){$b$}
\put(50,-14){$\sgl$}
\put(105,-14){$\sgl$}
\put(75,-12){$q$}
\put(75,20){$\tilde q$}
\put(44,32){$\sbottom$}
\end{picture} \\
\begin{picture}(200,70)(-10,0)
\DashLine(0,0)(100,0){5}
\GOval(50,0)(10,20)(0){0.5}
\DashLine(130,0)(230,0){5}
\GlueArc(180,0)(20,0,180){4}{5}
\DashLine(260,0)(360,0){5}
\ArrowLine(290,0)(330,0)
\CArc(310,0)(20,0,180)
\GlueArc(310,0)(20,0,180){4}{5}
\DashLine(390,0)(490,0){5}
\DashCArc(440,20)(20,0,360){5}
\Vertex(440,0){2}
\put(88,-3){$=$}
\put(193,-3){$+$}
\put(297,-3){$+$}
\put(10,6){$\sbottom$}
\put(70,6){$\sbottom$}
\put(113,6){$\sbottom$}
\put(143,-13){$\sbottom$}
\put(143,25){$g$}
\put(175,6){$\sbottom$}
\put(218,6){$\sbottom$}
\put(248,-12){$b$}
\put(248,25){$\sgl$}
\put(280,6){$\sbottom$}
\put(323,6){$\sbottom$}
\put(351,35){$\sbottom$}
\put(383,6){$\sbottom$}
\end{picture} \\
\caption{\it Two-loop diagrams of (a) the SUSY--QCD and (b) the
top-induced mixed SUSY--QCD/electroweak contributions to the bottom
self-energy involving bottom quarks $b$, bottom $\sbottom$ and top $\st$
squarks, gluons $g$, gluinos $\sgl$ and charged Higgsinos $\tilde
h^\pm$. The squark-quark contributions to the gluino propagator have to
be summed over all quark/squark flavors $q/\tilde q$ including both
directions of the flavor flow due to the Majorana nature of the gluino.}
\label{fig:2loop_examples}
\end{figure}
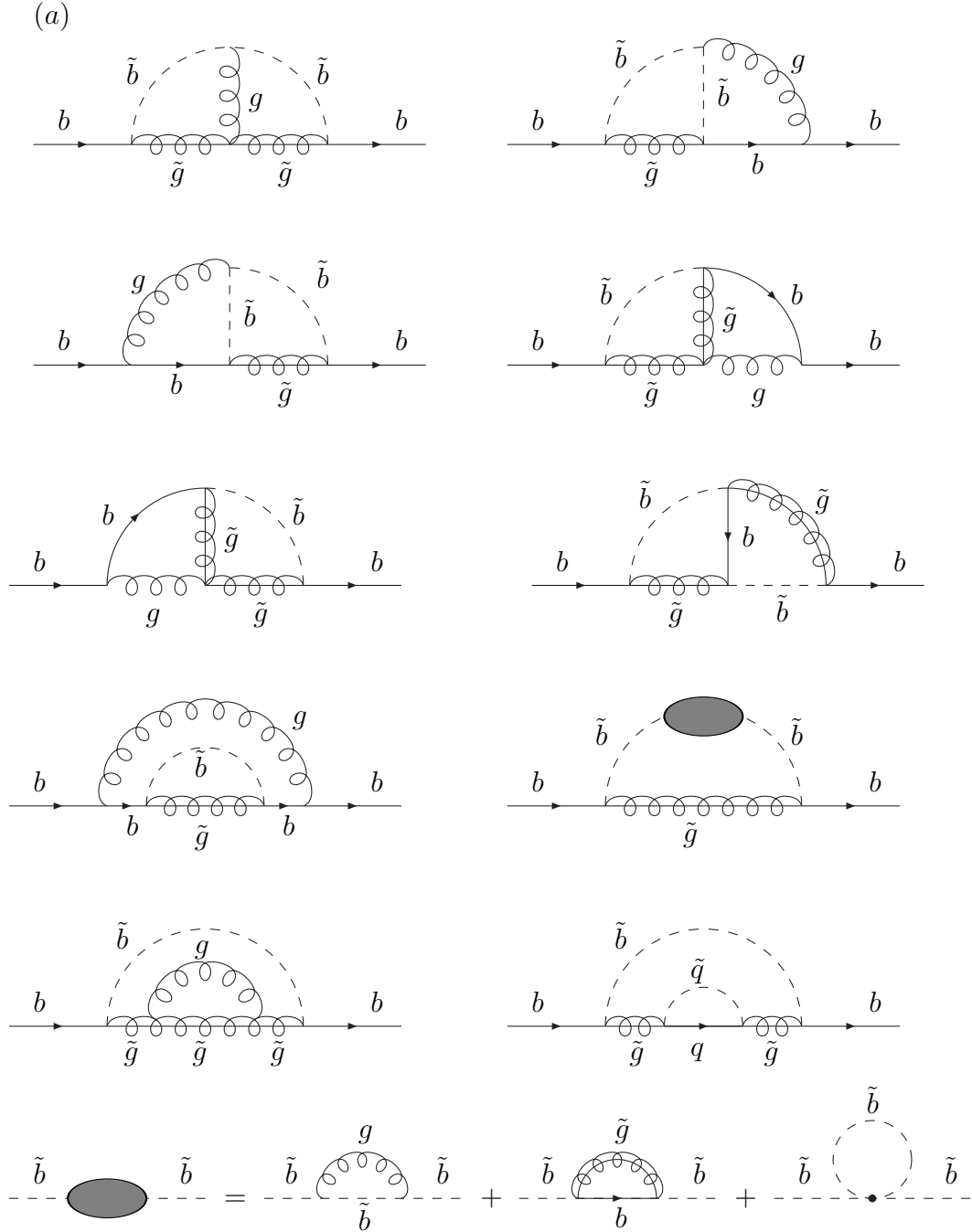
\addtocounter{figure}{-1}
\begin{figure}[hbt]
\SetScale{0.8}
\begin{picture}(200,90)(-20,0)
\ArrowLine(0,0)(50,0) 
\Line(50,0)(100,0) 
\Photon(50,0)(100,0){5}{3}
\ArrowLine(100,0)(150,0)
\DashLine(100,50)(100,0){5}
\ArrowLine(150,0)(200,0) 
\GlueArc(100,0)(50,0,90){5}{5}
\DashCArc(100,0)(50,90,180){5}
\put(10,6){$b$}
\put(148,6){$b$}
\put(86,16){$\st$}
\put(56,-16){$\tilde h^\pm$}
\put(100,-12){$b$}
\put(44,32){$\st$}
\put(116,32){$g$}
\put(0,50){$(b)$}
\end{picture}
\begin{picture}(200,90)(-20,0)
\ArrowLine(0,0)(50,0) 
\ArrowLine(50,0)(100,0) 
\Line(100,0)(150,0)
\Photon(100,0)(150,0){5}{3}
\DashLine(100,50)(100,0){5}
\ArrowLine(150,0)(200,0) 
\GlueArc(100,0)(50,90,180){5}{5}
\DashCArc(100,0)(50,0,90){5}
\put(10,6){$b$}
\put(148,6){$b$}
\put(86,16){$\st$}
\put(56,-12){$b$}
\put(100,-15){$\tilde h^\pm$}
\put(40,32){$g$}
\put(116,32){$\st$}
\end{picture} \\
\begin{picture}(200,90)(-20,0)
\ArrowLine(0,0)(50,0) 
\Line(50,0)(100,0) 
\Photon(50,0)(100,0){5}{3}
\DashLine(100,0)(150,0){5}
\ArrowLine(100,50)(100,0)
\ArrowLine(150,0)(200,0) 
\CArc(100,0)(50,0,90)
\GlueArc(100,0)(50,0,90){5}{5}
\DashCArc(100,0)(50,90,180){5}
\put(10,6){$b$}
\put(148,6){$b$}
\put(86,16){$t$}
\put(56,-15){$\tilde h^\pm$}
\put(100,-15){$\sbottom$}
\put(44,32){$\st$}
\put(116,32){$g$}
\end{picture}
\begin{picture}(200,90)(-20,0)
\ArrowLine(0,0)(50,0) 
\Line(50,0)(100,0) 
\Gluon(50,0)(100,0){5}{3}
\DashLine(100,0)(150,0){5}
\ArrowLine(100,50)(100,0)
\ArrowLine(150,0)(200,0) 
\CArc(100,0)(50,0,90)
\PhotonArc(100,0)(50,0,90){5}{5}
\DashCArc(100,0)(50,90,180){5}
\put(10,6){$b$}
\put(148,6){$b$}
\put(86,16){$t$}
\put(56,-15){$\sgl$}
\put(100,-15){$\st$}
\put(44,32){$\sbottom$}
\put(116,32){$\tilde h^\pm$}
\end{picture} \\
\begin{picture}(200,90)(-10,0)
\ArrowLine(0,0)(50,0) 
\ArrowLine(150,0)(200,0) 
\ArrowLine(50,0)(70,0) 
\ArrowLine(130,0)(150,0) 
\Line(70,0)(130,0) 
\Photon(70,0)(130,0){5}{4}
\GlueArc(100,0)(50,0,180){5}{10}
\DashCArc(100,0)(30,0,180){5}
\put(10,6){$b$}
\put(148,6){$b$}
\put(48,-12){$b$}
\put(76,-16){$\tilde h^\pm$}
\put(112,-12){$b$}
\put(76,12){$\st$}
\put(116,34){$g$}
\end{picture}
\begin{picture}(200,90)(-10,0)
\ArrowLine(0,0)(50,0)
\Line(50,0)(150,0)
\Photon(50,0)(150,0){5}{7}
\ArrowLine(150,0)(200,0)
\DashCArc(100,0)(50,0,180){5}
\GOval(100,46)(10,20)(0){0.5}
\put(10,6){$b$}
\put(148,6){$b$}
\put(72,-15){$\sgl$}
\put(36,25){$\st$}
\put(116,25){$\st$}
\end{picture} \\
\begin{picture}(200,70)(-10,0)
\DashLine(0,0)(100,0){5}
\GOval(50,0)(10,20)(0){0.5}
\DashLine(130,0)(230,0){5}
\GlueArc(180,0)(20,0,180){4}{5}
\DashLine(260,0)(360,0){5}
\ArrowLine(290,0)(330,0)
\CArc(310,0)(20,0,180)
\GlueArc(310,0)(20,0,180){4}{5}
\DashLine(390,0)(490,0){5}
\DashCArc(440,20)(20,0,360){5}
\Vertex(440,0){2}
\put(88,-3){$=$}
\put(193,-3){$+$}
\put(297,-3){$+$}
\put(10,6){$\st$}
\put(70,6){$\st$}
\put(113,6){$\st$}
\put(143,-13){$\st$}
\put(143,25){$g$}
\put(175,6){$\st$}
\put(218,6){$\st$}
\put(248,-12){$t$}
\put(248,25){$\sgl$}
\put(280,6){$\st$}
\put(323,6){$\st$}
\put(351,35){$\st$}
\put(383,6){$\st$}
\end{picture} \\
\caption{\it Cont'd.}
\end{figure}
We have computed the supersymmetric QCD corrections to the effective
bottom Yukawa couplings of Eq.~(\ref{eq:rescoup}), i.e.~the SUSY--QCD
corrected effective Lagrangian of Eq.(\ref{eq:leff}) at NNLO. The
two-loop calculation split up into the ${\cal O}(\alpha_s^2\mu\tgb)$
corrections to $\Delta_b^{QCD(1)}$ and the ${\cal
O}(\alpha_s\lambda_t^2 A_t\tgb)$ corrections to $\Delta_b^{elw(1)}$.
The relevant two-loop diagrams contributing to the bottom
self-energy are shown in Figure \ref{fig:2loop_examples}. For the
determination of the NNLO corrections single scalar mass insertions in
each bottom- and top-squark propagator analogous to the one-loop level
have to be included and added in all possible ways. This means that
e.g.~the first diagram leads to two contributions, one with the mass
insertion in the left sbottom propagator and the other with the mass
insertion in the right sbottom propagator as depicted in
Fig.~\ref{fg:insertions}.
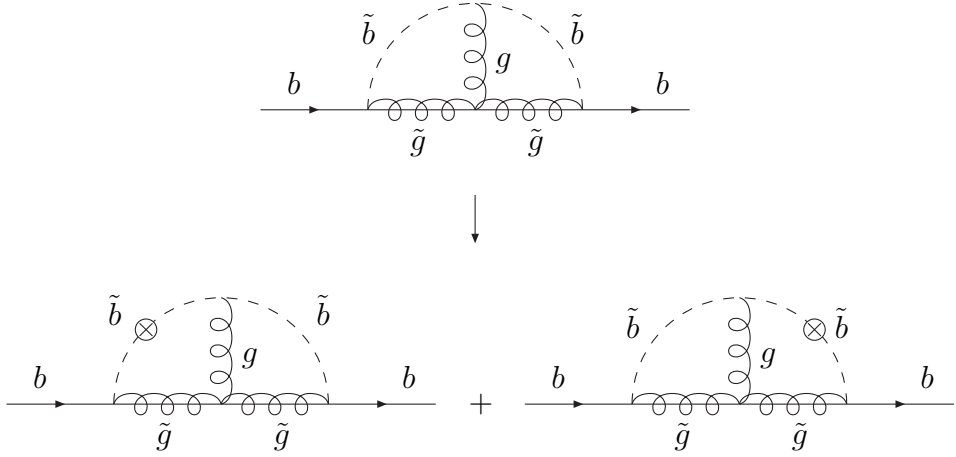
\begin{figure}[hbtp]
\SetScale{0.8}
\begin{picture}(200,60)(-115,0)
\ArrowLine(0,0)(50,0) 
\Line(50,0)(150,0) 
\Gluon(50,0)(100,0){5}{3}
\Gluon(100,0)(150,0){5}{3}
\Gluon(100,50)(100,0){5}{3}
\ArrowLine(150,0)(200,0) 
\DashCArc(100,0)(50,0,180){5}
\Line(100,-40)(100,-60) 
\ArrowLine(100,-60)(100,-61) 
\put(10,6){$b$}
\put(148,6){$b$}
\put(88,16){$g$}
\put(56,-15){$\sgl$}
\put(100,-15){$\sgl$}
\put(38,25){$\sbottom$}
\put(116,25){$\sbottom$}
\end{picture} \\
\begin{picture}(200,110)(-20,0)
\ArrowLine(0,0)(50,0) 
\Line(50,0)(150,0) 
\Gluon(50,0)(100,0){5}{3}
\Gluon(100,0)(150,0){5}{3}
\Gluon(100,50)(100,0){5}{3}
\ArrowLine(150,0)(200,0) 
\DashCArc(100,0)(50,0,180){5}
\CrossedCircle{65}{35}{62}{32}{68}{38}
\put(10,6){$b$}
\put(148,6){$b$}
\put(88,16){$g$}
\put(56,-15){$\sgl$}
\put(100,-15){$\sgl$}
\put(38,30){$\sbottom$}
\put(116,30){$\sbottom$}
\put(173,-3){+}
\end{picture}
\begin{picture}(200,110)(-10,0)
\ArrowLine(0,0)(50,0) 
\Line(50,0)(150,0) 
\Gluon(50,0)(100,0){5}{3}
\Gluon(100,0)(150,0){5}{3}
\Gluon(100,50)(100,0){5}{3}
\ArrowLine(150,0)(200,0) 
\DashCArc(100,0)(50,0,180){5}
\CrossedCircle{135}{35}{132}{32}{138}{38}
\put(10,6){$b$}
\put(148,6){$b$}
\put(88,16){$g$}
\put(56,-15){$\sgl$}
\put(100,-15){$\sgl$}
\put(38,25){$\sbottom$}
\put(116,25){$\sbottom$}
\end{picture} \\
\caption{\it All possible mass insertions into the first diagram of
Fig.~\ref{fig:2loop_examples} contributing to the leading terms of
$\Delta_b$ at NNLO.}
\label{fg:insertions}
\end{figure}
We have performed the whole calculation in dimensional regularization.
The bottom quark momentum and its mass have been put to zero while
keeping the bottom Yukawa coupling $\lambda_b$ finite in the mass
insertions and the charged Higgsino couplings. All supersymmetric
particles as well as the top quark have been treated with full mass
dependence\footnote{We have neglected the tiny contributions of ${\cal
O}(m_t^2 (A_t-\mu/\tgb)/M_{SUSY}^3)$ of the off-diagonal stop
propagators in the gluino self-energy of the last two-loop diagram (for
$q = t, \tilde q=\st$) of Fig.~\ref{fig:2loop_examples}a and in the
gluino mass counter term $\delta \Mg$ in Eq.~(\ref{eq:ct}).}. Since the
two-loop contributions are computed in the limit of vanishing external
momenta the calculation requires the determination of the leading
two-loop vacuum integrals in the heavy mass expansion.  These integrals
can be expressed in terms of the one-loop one-point integral $\Ao{m}$
\cite{passvel}
\beq
A_0(m) = \bar \mu^{4-n} \int \frac{d^nk}{(2\pi)^n} \frac{1}{k^2-m^2}
\label{eq:a0}
\eeq
where $\bar \mu$ denotes the 't Hooft mass, and the two-loop master
integral $T_{134}(m_1,m_3,m_4)$ \cite{t134}
\beq
T_{134}(m_1,m_3,m_4) = \bar \mu^{2(4-n)} \int \frac{d^nk}{(2\pi)^n}
\frac{d^nq}{(2\pi)^n} \frac{1}{(k^2-m_1^2)[(k-q)^2-m_3^2](q^2-m_4^2)}
\eeq
where we work in $n=4-2\epsilon$ dimensions. Integrals with higher
powers of the corresponding propagators can be reduced to the basic
integrals $A_0$ and $T_{134}$ by means of integration-by-parts methods
\cite{ibpmethods}. In this way all singularities have been separated
from the finite contributions analytically as poles in the parameter
$\epsilon$.

\subsection{Renormalization}
Contrary to the finite one-loop results, the two-loop corrections
$\Delta_b^{(2)}$ are UV-divergent. A finite and $\bar\mu$-independent
result is obtained after renormalization of the masses and couplings
contributing to the one-loop result.  The heavy masses $m_{\sbottom_i},
m_{\st_i}, m_{\sgl}$ appearing in the propagators have been renormalized
on-shell. The trilinear coupling $A_t$ has been treated in the on-shell
scheme, too, where its counter term is derived from the counter terms of
the stop mixing angle and masses, and the top Yukawa
coupling\footnote{Note that in the desired order the mixing angle
counter term does not depend on the momentum of the off-diagonal
self-energy contribution, since the gluino contribution does not
contribute to our perturbative order, $\delta\theta_t = iC_F g_s^2 \sin
2\theta_t \cos 2\theta_t
[A_0(\Mt{2})-A_0(\Mt{1})]/(\Mt{2}^2-\Mt{1}^2)$.},
\beq
\delta A_t = \frac{\Mt{1}^2 - \Mt{2}^2}{2m_t} \left[ 2\cos 2\theta_t
\delta\theta_t - \sin 2\theta_t \frac{\delta m_t}{m_t} \right] +
\frac{\sin 2\theta_t}{2m_t} \left[ \delta \Mt{1}^2 - \delta \Mt{2}^2
\right]
\eeq
where $\delta m_t$ denotes the on-shell counter term of the top mass,
\beq
\frac{\delta m_t}{m_t} = iC_F g_s^2 \left\{ \frac{A_0(m_t)}{m_t^2}
+ 2B_0(m_t^2;0,m_t) + B_1(m_t^2;\Mg,m_{\st_1}) +
B_1(m_t^2;\Mg,m_{\st_2}) \right\}
\eeq
The strong coupling $\alpha_s$ and the top Yukawa coupling $\lambda_t$
have been defined in the $\MS$ scheme with 5 active flavors, i.e.~the
top quark and the supersymmetric particles have been decoupled from the
scale dependence of the strong coupling $\alpha_s(\mu_R)$. Care has to
be taken to include only the desired order, i.e.~${\cal O}(\alpha_s^2
\mu \tgb/M_{SUSY})$ for $\Dmb^{QCD}$ and ${\cal O}(\alpha_s \lambda^2_t
A_t \tgb/M_{SUSY})$ for $\Dmb^{elw}$. In this order the trilinear
coupling $A_t$ only receives a finite renormalization at NLO.

The complete set of counter terms is given by
\bea
\Mq{i}^{0\,2} & = & \Mq{i}^2 + \dMq{i}^2 \nonumber \\
\Mg^0 & = & \Mg + \delta \Mg \nonumber \\
\lambda_t^0 & = & \lambda_t(\mu_R) + \delta \lambda_t \nonumber \\
\alpha_s^0 & = & \alpha_s(\mu_R) +\delta \alpha_s \nonumber \\
A_t^0 & = & A_t + \delta A_t
\eea
with the following explicit expressions ($C_A=3$)
\bea
\delta \Mt{i}^{2} & = & -iC_F g_s^2 \left\{ 2A_0(\Mt{i}) - 2A_0(\Mg)
- 2A_0(m_t) - 4\Mt{i}^2 B_0(\Mt{i}^2;0,\Mt{i}) \right. \nonumber \\
& & \hspace*{2cm} \left. +2(\Mt{i}^2-\Mg^2-m_t^2)B_0(\Mt{i}^2;\Mg,m_t) \right\} \nonumber \\
\delta \Mb{i}^{2} & = & -iC_F g_s^2 \left\{ 2 A_0(\Mb{i}) -2A_0(\Mg)  -
4\Mb{i}^2 B_0(\Mb{i}^2;0,\Mb{i}) \right. \nonumber \\
& & \hspace*{2cm} \left. +2(\Mb{i}^2-\Mg^2)B_0(\Mb{i}^2;\Mg,0)
\right\} \nonumber \\
\frac{\delta \Mg}{\Mg} & = & i g_s^2 \left\{C_A \left[ 2 B_0(\Mg^2;0,\Mg) -
2B_1(\Mg^2;0,\Mg) -\frac{i}{(4\pi)^2} \right] \phantom{\frac{1}{2}}
\right. \nonumber \\
& & \left. \hspace*{1.3cm} - \sum_{q}\left[
B_0(\Mg^2;m_{\squark_1},m_q) + B_0(\Mg^2;m_{\squark_2},m_q) \right.
\right. \nonumber \\
& & \left. \hspace*{2cm} \phantom{\frac{1}{2}} \left.
+ B_1(\Mg^2;m_{\squark_1},m_q) + B_1(\Mg^2;m_{\squark_2},m_q) \right]
\right\} \nonumber \\
\frac{\delta\lambda_t}{\lambda_t} & = & C_F\frac{\alpha_s}{\pi}
\Gamma(1+\epsilon) (4\pi)^\epsilon \frac{3}{4} \left\{-\frac{1}{\epsilon} +\log
\frac{\mu_R^2}{\bar \mu^2}\right\}
+iC_F g_s^2 \left\{ B_1(m_t^2;\Mg,\Mt{1}) +
B_1(m_t^2;\Mg,\Mt{2}) \right\} \nonumber \\
\frac{\delta\alpha_s}{\alpha_s} & = &
\frac{\alpha_s}{\pi}\Gamma(1+\epsilon) (4\pi)^\epsilon \left\{
\left(-\frac{1}{\epsilon}+\log\frac{\mu_R^2}{\bar \mu^2}\right) \beta_0 +
\frac{C_A}{6}\log\frac{\mu_R^2}{\Mg^2} +
\frac{1}{6}\log\frac{\mu_R^2}{m_t^2} \right. \nonumber \\
& & \hspace*{3cm} \left. + \sum_{\tilde q_i} \frac{1}{24}
\log\frac{\mu_R^2}{\Mq{i}^2} \right\} \nonumber \\
\frac{\delta A_t}{A_t} & = & -iC_F g_s^2 \left\{ \frac{A_0(m_t)}{m_t^2}
+ 2B_0(m_t^2;0,m_t) + B_1(m_t^2;\Mg,m_{\st_1}) + B_1(m_t^2;\Mg,m_{\st_2})
\right. \nonumber \\
& & \left.
- 4\frac{m_{\st_1}^2B_0(m_{\st_1}^2;0,m_{\st_1}) -
  m_{\st_2}^2B_0(m_{\st_2}^2;0,m_{\st_2})}{m_{\st_1}^2-m_{\st_2}^2}
\right. \nonumber \\
& & \left.
+2\frac{(m_{\st_1}^2-\Mg^2-m_t^2)B_0(m_{\st_1}^2;\Mg,m_t) -
(m_{\st_2}^2-\Mg^2-m_t^2)B_0(m_{\st_2}^2;\Mg,m_t)}{m_{\st_1}^2-m_{\st_2}^2}
\right\}
\label{eq:ct}
\eea
where the lowest order beta function is given by
\beq
\beta_0 = \frac{3C_A-N_F-1}{4}
\eeq
with $N_F=5$ light flavors ($m_q=0$) and summing over $N_F+1=6$
quark/squark flavors in the expressions above. The masses of all quarks
except the top have been put to zero. In addition to the one-loop scalar
integral $A_0$ of Eq.(\ref{eq:a0}) the residual scalar one-loop
integrals are defined as \cite{passvel}
\bea
B_0(p^2;m_1,m_2) & = & \bar \mu^{4-n} \int \frac{d^nk}{(2\pi)^n}
\frac{1}{(k^2-m_1^2) [(k+p)^2-m_2^2]} \nonumber \\
B_1(p^2;m_1,m_2) & = & \frac{A_0(m_1)-A_0(m_2)
-(p^2+m_1^2-m_2^2)B_0(p^2;m_1,m_2)}{2p^2}
\eea
A few comments are in order. Since we calculate the ${\cal O}(\alpha_s^2
\mu \tgb)$ and ${\cal O}(\alpha_s \lambda_t^2 A_t \tgb)$ corrections we
only included contributions to these orders in our counter terms. This
plays a role for the counter terms of the gluino mass, the stop masses,
the top Yukawa coupling $\lambda_t$ and the trilinear coupling $A_t$.
In the gluino mass counter term we have neglected contributions of the
order $m_t/\Mg$, in the counter terms of the stop masses and the top
Yukawa coupling $\lambda_t$ we deleted terms of ${\cal O}(m_t A_t)$, and
finally in the counter term of the trilinear coupling $A_t$ we only
included contributions linear in $A_t$.  Moreover, we have put the
bottom mass $m_b$ to zero everywhere apart from the Yukawa coupling
$\lambda_b$ in the mass insertions and the Higgsino couplings. This
leads to the simplified counter terms listed above \cite{noth}.

The counter term for the strong coupling constant contains the
logarithmic finite terms required for the decoupling of the top quark
and the supersymmetric particles from its scale dependence, i.e. the
coupling $\alpha_s$ runs with only five light flavors,
\beq
\mu_R^2 \frac{\partial \alpha_s}{\partial \mu_R^2} = -\beta_0^L
\frac{\alpha_s^2}{\pi} + {\cal O}(\alpha_s^3)
\eeq
with ($N_F=5$)
\beq
\beta_0^L = \frac{11C_A-2N_F}{12}
\eeq

\paragraph{\it Anomalous Counter Terms.}
A complication occurs in supersymmetric theories if dimensional
regularization is employed, since in $n\neq4$ dimensions the gluons
acquire $n-2$ degrees of freedom, while the gluinos still possess 2
degrees of freedom. Thus a mismatch between the gluons and their
supersymmetric partners is introduced at ${\cal O}(\epsilon)$ which
violates the supersymmetric relations between related couplings and
masses. In order to restore supersymmetry at the renormalized level
finite anomalous counter terms have to be introduced which, however, are
uniquely fixed by the corresponding supersymmetric identities
\cite{anom}.

In our calculation this mismatch of degrees of freedom between
supersymmetric partners induces finite differences between the Yukawa
coupling $\hat g_s$ of quarks, squarks and gluinos and the corresponding
gauge coupling $g_s$,
\beq
\hat{g}_s = \gs \Big[1 + \frac{\as}{2\pi}\Big(\frac{C_A}{3} -
\frac{C_F}{4}\Big)\Big]
\eeq
and between the quark Yukawa couplings (generically denoted by
$\lambda_{Hqq}$), the corresponding Higgs couplings to squarks
(generically $\lambda_{H\squark\squark}$) and the Higgsino Yukawa
couplings to quarks and squarks (generically $\lambda_{\tilde H q
\squark}$),
\beq
\lambda_{Hqq} = \lambda_{H\tilde{q}\tilde{q}} \Big[ 1 + \frac{C_F}{4}
\frac{\as}{\pi} \Big] = \lambda_{\tilde{H}q\tilde{q}} \Big[ 1 +
\frac{3}{8}C_F \frac{\as}{\pi} \Big]
\eeq
In our calculation we have expressed all couplings in terms of the
Standard Model $\overline{MS}$ Yukawa coupling $\lambda_{Hqq}$. The
anomalous counter terms for the $H\squark\squark$ vertex has already
been included in the counter term $\delta A_t$ given above. The
left-over anomalous counter terms are those for the strong coupling
$\hat g_s$ arising in the SUSY--QCD corrections $\Delta_b^{QCD (1)}$
at the one-loop level, the bottom Yukawa coupling $\lambda_b$ of the
mass insertions of $\Delta_b^{QCD (1)}$, which corresponds to the
generic coupling $\lambda_{H\squark\squark}$, the bottom Yukawa coupling
$\lambda_b$ corresponding to $\lambda_{\tilde{H}q\tilde{q}}$ generically
in the SUSY-electroweak corrections $\Delta_b^{elw (1)}$ and the top
Yukawa coupling factor $\lambda_t^2$ of $\Delta_b^{elw (1)}$, which
corresponds to the generic product $\lambda_{H\tilde{q}\tilde{q}}
\lambda_{\tilde{H}q\tilde{q}}$. This results in the following total
anomalous counter terms \cite{noth}
\bea
\delta \Delta_{b;anom}^{QCD} & = & \left(\frac{C_A}{3}- \frac{C_F}{4}
- \frac{C_F}{4} \right) \frac{\alpha_s}{\pi} \Dmb^{QCD (1)} =
\left(\frac{C_A}{3}- \frac{C_F}{2}\right) \frac{\alpha_s}{\pi} \Dmb^{QCD
(1)} \nonumber \\
\delta \Delta_{b;anom}^{elw} & = & \left(-\frac{C_F}{4} -
\frac{3}{8}C_F - \frac{3}{8}C_F \right) \frac{\alpha_s}{\pi} \Dmb^{elw
(1)} = -C_F \frac{\alpha_s}{\pi} \Dmb^{elw (1)}
\eea
These counter terms have been added to the final results after including
the counter terms of Eq.(\ref{eq:ct}) to obtain consistent final results.

\subsection{Resummation}
According to the power-counting arguments of Ref.~\cite{GHS} the leading
NLO contributions of ${\cal O}(\mu \tgb)$ can only emerge from {\it
single} mass insertions as depicted in
Fig.~\ref{fig:1loop_Selfenergies}, i.e.~NNLO terms of ${\cal O}(\mu^2
{\rm tg}^2\beta)$ are absent, since a second mass insertion as required
by the second power in $\mu\tgb$ is suppressed by an additional power of
$m_b/M_{SUSY}$. In complete analogy the absence of ${\cal O}(A_t^2 {\rm
tg}^2\beta)$ in the SUSY-electroweak part $\Delta_b^{elw}$ can be
proven. Thus the leading two-loop corrections are of ${\cal
O}(\alpha_s^2\mu \tgb)$ and ${\cal O}(\alpha_s\lambda^2_t A_t \tgb)$
respectively, arising from {\it single} mass insertions in the two-loop
bottom self-energy diagrams contributing to $\Sigma_S(m_b)$ in
Eq.~(\ref{eq:sigma_s}). These are indeed all contributions taken into
account in this work. At higher orders terms of ${\cal O}(\alpha_s \mu^2
{\rm tg}^2\beta)$ and ${\cal O}(\alpha_s A_t^2 {\rm tg}^2\beta)$ are
absent so that our results of ${\cal O}(\alpha_s^2\mu \tgb)$ and ${\cal
O}(\alpha_s\lambda^2_t A_t \tgb)$ are automatically resummed if included
in $\Delta_b$ according to Eq.~(\ref{eq:leff}) with
\bea
\Delta_b & = & \Delta_b^{QCD} + \Delta_b^{elw} \nonumber \\
\Delta_b^{QCD} & = & \Delta_b^{QCD\,(1)} + \Delta_b^{QCD\,(2)}
\nonumber \\
\Delta_b^{elw} & = & \Delta_b^{elw\,(1)} + \Delta_b^{elw\,(2)}
\eea
where $\Delta_b^{QCD\,(2)}$ and $\Delta_b^{elw\,(2)}$ denote our novel
NNLO corrections. These expressions thus resum all terms of ${\cal
O}(\alpha_s^n\mu^n{\rm tg}^n\beta)$, ${\cal O}(\alpha_s^{n+1}\mu^n{\rm
tg}^n\beta)$, ${\cal O}(\lambda_t^{2n}A_t^n{\rm tg}^n\beta)$ and ${\cal
O}(\alpha_s \lambda_t^{2n}A_t^n{\rm tg}^n\beta)$.

\subsection{Limits}
The final results for the two-loop corrections $\Delta_b^{QCD\,(2)}$ and
$\Delta_b^{elw\,(2)}$ are too lengthy to be displayed here. However,
they can be given explicitly in certain limits: \\[0.5cm]
{\it a)} $m_t^2 \ll \Mg^2 = \Mb{i}^2 = \mu^2 = \Mt{i}^2 \equiv
M^2$:\footnote{It should be noted that degenerate squark masses require
$A_b = \mu\tgb$ and $A_t = \mu / \tgb$. Moreover, due to the different
$D$-term contributions to the $SU(2)$-relation between the left-handed
stop and sbottom mass terms the equality of the stop and sbottom masses
cannot be realized. However, we display the full result in this limit as
a good approximation for small mass splittings.}
\bea
\Delta_b^{QCD\,(1)} & = & \frac{C_F}{4} \frac{\alpha_s(\mu_R)}{\pi} \tgb
\nonumber \\
\Delta_b^{QCD\,(2)} & = & \frac{\alpha_s}{\pi}\left\{\frac{C_A}{3} +
C_F +\frac{N_F+1}{4} +\frac{1}{6}\log\frac{M^2}{m_t^2} + \beta_0^L
\log\frac{\mu_R^2}{M^2} \right\} \Delta_b^{QCD\,(1)} \nonumber \\
\Delta_b^{elw\,(1)} & = & \frac{1}{2} \frac{\lambda_t^2(\mu_R)}{(4\pi)^2}
\frac{A_t\tgb}{M} \nonumber \\
\Delta_b^{elw\,(2)} & = & C_F \frac{\alpha_s}{\pi}\left\{ \frac{7}{4} +
\frac{3}{2} \log\frac{\mu_R^2}{m_t M} \right\} \Delta_b^{elw\,(1)}
\eea
where terms of ${\cal O}(m_t^2/M^2)$ have been neglected. The
logarithmic top mass term of $\Delta_b^{QCD\,(2)}$ can be absorbed in
the running strong coupling by evolving it with 6 active flavors.
However, since the top mass is not much lighter than the supersymmetric
masses this logarithm does not become that large that this change of
scheme is required. The last logarithm compensates the scale dependence
of the strong coupling $\alpha_s(\mu_R)$ in the one-loop expression
$\Delta_b^{QCD\,(1)}$. We have verified that our results for
$\Delta_b^{QCD\,(1,2)}$ can be obtained from the $X_b$ terms of
Eq.~(4.8) in Ref.~\cite{mihaila} if the scheme transformations and
decoupling relations of the strong coupling $\alpha_s$ and the trilinear
coupling $A_b$ are taken into account properly. Moreover, we have found
agreement with the $a_q$ terms of Eq.~(61) in the first paper of
Ref.~\cite{bednyakov} after appropriate scheme transformations of
$\alpha_s, m_{\tilde g}, m_{\tilde b_{1/2}}$ and the bottom mass and the
necessary subtraction of the product of the vectorial part of the
one-loop self-energy $\Sigma_V(m_b)$ and the scalar part $\Sigma_S(m_b)$
which contributes to the relation between the pole quark mass and the
$\overline{DR}$ mass in addition to the pure self-energy at the two-loop
level.

The final expressions for $\Delta_b^{elw\,(2)}$ exhibits a logarithmic
term of the common supersymmetric mass $M$ which can be absorbed in the
low-energy Yukawa couplings of the charged Higgsinos to stops and bottom
quarks which have to be defined with properly decoupled supersymmetric
contributions. These low-energy Yukawa couplings have to be evaluated at
scales of the order of the top quark mass. This, however, is only
necessary if the mass splitting between the top quark and the
supersymmetric particles becomes large, giving rise to logarithms of
their mass ratios. This does not occur in realistic scenarios so that we
did not perform this low-energy decoupling. \\[0.5cm]
\noindent
{\it b)} $m_t^2, \Mb{i}^2, \mu^2, \Mt{i}^2 \ll \Mg^2$:
\bea
\Delta_b^{QCD\,(1)} & = & \frac{C_F}{2} \frac{\alpha_s(\mu_R)}{\pi} \frac{\mu
\tgb}{\Mg}~\left\{\log\frac{\Mg^2}{\Mb{2}^2}
-\frac{\Mb{1}^2}{\Mb{2}^2-\Mb{1}^2} \log\frac{\Mb{2}^2}{\Mb{1}^2}
\right\} \nonumber \\
\Delta_b^{QCD\,(2)} & = & \frac{\alpha_s}{\pi}\left\{\frac{4}{3}C_A +
C_F\left[ \log\frac{\Mg^2}{\Mb{2}^2} -\frac{\Mb{1}^2}{\Mb{2}^2-\Mb{1}^2}
\log\frac{\Mb{2}^2}{\Mb{1}^2} + \frac{5}{2} \right] -\frac{N_F+1}{2} \right.
\nonumber \\
& & \left. \qquad +\frac{1}{6}\log\frac{\Mg^2}{m_t^2} +
\frac{1}{24}\sum_{\tilde q} \log\frac{\Mg^2}{m_{\tilde q}^2} + \beta_0^L
\log\frac{\mu_R^2}{\Mg^2} \right\} \Delta_b^{QCD\,(1)} \nonumber \\
\Delta_b^{elw\,(2)} & = & C_F \frac{\alpha_s}{\pi} \left\{ \frac{23}{8}
+ \frac{3}{2} \log\frac{\mu_R^2}{m_t \Mg} \right\} \Delta_b^{elw\,(1)}
\eea
The next non-leading term in this limit is of ${\cal O}(\log^{-1}
\Mg^2/\Mb{i}^2)$ for the SUSY--QCD part $\Delta_b^{QCD\,(2)}$ and of
${\cal O}(\max \{\mu^2, \Mt{i}^2\}/\Mg^2)$ for the SUSY--electroweak
part $\Delta_b^{elw\,(2)}$.
The expression of $\Delta_b^{elw\,(1)}$ is not altered from
Eq.~(\ref{eq:effpar}) in this limit so that we do not display it again.
The logarithms of the top and squark masses in $\Delta_b^{QCD\,(2)}$ can
be absorbed in the running QCD coupling $\alpha_s$ if the latter
includes the top quark and all squarks in its $\beta$ function. Since
the real mass splittings are not so large that these logarithms become
sizeable this procedure is not necessary. Again we have verified that
our results for $\Delta_b^{QCD\,(1,2)}$ in this limit can be derived
from the $X_b$ terms of Eq.~(4.10) in Ref.~\cite{mihaila} by taking into
account the decoupling properties and scheme transformations of
$\alpha_s$ and $A_b$ consistently and identifying all squark masses. As
for the previous limit the logarithm of the gluino mass in
$\Delta_b^{elw\,(2)}$ can be absorbed in the low-energy Yukawa couplings
of the Higgsinos to stops and bottom quarks by decoupling the gluino
contributions consistently. Since these logarithms do not become large
in realistic cases we did not perform this decoupling but kept the full
supersymmetric relations to the related Yukawa and gauge couplings.

\section{Results}
The results obtained in this work have been inserted in the program
HDECAY \cite{hdecay} which calculates the MSSM Higgs masses and
couplings according to Ref.~\cite{rgi} as well as all partial decay
widths and branching ratios including the relevant higher-order
corrections \cite{spira:98}. For large values of $\tgb$ the decays of
the neutral Higgs bosons are dominated by the decays into $b\bar b$ and
$\tau^+\tau^-$.  Their branching ratios had been studied with the
one-loop expressions of the correction $\Delta_b$ of
Eq.~(\ref{eq:effpar}) in Ref.~\cite{GHS}.  It has been demonstrated that
the scale dependence of the strong coupling $\alpha_s(\mu_R)$ of
Eq.~(\ref{eq:effpar}) implies a theoretical uncertainty up to the 10\%
level for scenarios of large SUSY-QCD corrections to the bottom Yukawa
couplings such as, e.g., the `small $\alpha_{eff}$' scenario \cite{bench}.

\subsection{Higgs Decays into Bottom Quark Pairs}
The partial decay widths of the neutral Higgs bosons $\Phi=h,H,A$ into
bottom quark pairs, including QCD and SUSY--QCD corrections, can be cast
into the form \cite{GHS}
\begin{equation}
\Gamma [\Phi \, \to \, b{\overline{b}}] =
\frac{3G_F M_\Phi }{4\sqrt{2}\pi} \overline{m}_b^2(M_\Phi)
\left[ \Delta_{\rm QCD} + \Delta_t^\Phi \right] \tilde g_b^\Phi \left[
\tilde g_b^\Phi + \Delta_{SQCD}^{rem} \right]
\label{eq:gambb}
\end{equation}
where regular quark mass effects are neglected. The large logarithmic
part of the QCD corrections has been absorbed in the running
$\overline{\rm MS}$ bottom quark mass $\overline{m}_b(M_\Phi)$ at the
scale of the corresponding Higgs mass $M_\Phi$. The QCD corrections
$\Delta_{\rm QCD}$ and the top quark induced contributions
$\Delta_t^\Phi$ read as \cite{drees}
\begin{eqnarray}
\Delta_{\rm QCD} & = & 1 + 5.67 \frac{\alpha_s (M_\Phi)}{\pi} + (35.94 -
1.36
N_F) \left( \frac{\alpha_s (M_\Phi)}{\pi} \right)^2 \nonumber \\
& & + (164.14 - 25.77 N_F + 0.259 N_F^2) \left(
\frac{\alpha_s(M_\Phi)}{\pi}
\right)^3 \nonumber \\
\Delta_t^{h/H} & =& \frac{g_t^{h/H}}{g_b^{h/H}}~\left(\frac{\alpha_s
(M_{h/H})}{\pi}
\right)^2 \left[ 1.57 - \frac{2}{3} \log \frac{M_{h/H}^2}{M_t^2}
+ \frac{1}{9} \log^2 \frac{\overline{m}_b^2
(M_{h/H})}{M_{h/H}^2}\right]\nonumber \\
\Delta_t^A & = & \frac{g_t^A}{g_b^A}~\left(\frac{\alpha_s (M_A)}{\pi}
\right)^2
\left[ 3.83 - \log \frac{M_A^2}{M_t^2} + \frac{1}{6} \log^2
\frac{\overline{m}_b^2 (M_A)}{M_A^2} \right]
\end{eqnarray}
where $N_F=5$ active flavors are taken into account. In the
intermediate and large Higgs mass regimes the QCD corrections reduce the
$b\bar b$ decay widths by about 50\% due to the large logarithmic
contributions.

The main part of the SUSY--QCD corrections \cite{solaeberl} has been
absorbed in the effective bottom Yukawa couplings $\tilde g_b^\phi$ of
Eq.~(\ref{eq:rescoup}). The remainder $\Delta_{SQCD}^{rem}$ is small in
phenomenologically relevant scenarios for large values of $\tgb$
\cite{GHS}. It should be noted that this remains true even for very
light SUSY masses in the range of ${\cal O}(100~{\rm GeV})$. This can be
understood by deriving the asymptotic expression of
$\Delta_{SQCD}^{rem}$ for large values of the Higgs masses compared to
the SUSY masses (we neglect the bottom mass here),
\beq
\Delta_{SQCD}^{rem} \to
C_F~\frac{\alpha_s(\mu_R)}{\pi}~m_{\sgl}~(A_b-\eta_\phi \mu\tgb)~
I(m^2_{\sbottom_1},m^2_{\sbottom_2},m^2_{\sgl})
\eeq
with the coefficients
\beq
\eta_H = \frac{\tga}{\tgb} \qquad \qquad \eta_A = -\frac{1}{{\rm
tg}^2\beta} \nonumber
\eeq
for the heavy scalar $H$ and the pseudoscalar $A$. Thus the large Higgs
mass limit approaches a finite expression of ${\cal O}(A_b/\mu\tgb)$
relative to the leading $\Delta_b$ terms for large values of $\tgb$.
This explains the validity of the $\Delta_b$ approximation even for
small SUSY masses compared to the Higgs masses. In the following we
include the mixed top Yukawa coupling induced SUSY--QCD/electroweak
corrections in the couplings $\tilde g_b^\phi$, too.

\subsection{Numerical Results}
The numerical analysis of the neutral Higgs boson decays into bottom
quark pairs is performed for two MSSM benchmark scenarios \cite{bench}
as representative cases\footnote{Note that in the small $\alpha_{eff}$
scenario we increased the gluino mass by a factor of two relative to
Ref.~\cite{bench} in order to enhance the size of $\Delta_b$. This value
for the gluino mass has also been used in Ref.~\cite{noth} so that
Eq.~(5) of that work has to be corrected. A general analysis of the
impact of the $\Delta_b$ terms in the MSSM is beyond the scope of our
work so that we restrict our numerical analysis to these two scenarios.}:
\begin{eqnarray}
\mbox{small $\alpha_{eff}$:} &&\tb = 30,\quad M_{\tilde Q} = 800~{\rm
GeV},\quad M_{\gluino} = 1000~{\rm GeV}, \nonumber \\
&&M_2 = 500~{\rm GeV},\quad A_b = A_t = -1.133~{\rm TeV}, \quad \mu =
2~{\rm TeV} \nonumber \\ \nonumber \\
\mbox{gluophobic:} &&\tb = 30,\quad M_{\tilde Q} = 350~{\rm GeV},\quad
M_{\gluino} = 500~{\rm GeV}, \nonumber \\
&&M_2 = 300~{\rm GeV},\quad A_b = A_t = -760~{\rm GeV},\quad \mu =
300~{\rm GeV}
\end{eqnarray}
We use the RG-improved two-loop expressions for the Higgs masses and
couplings of Ref.~\cite{rgi}. Thus the leading one- and two-loop
corrections have been included in the effective mixing angle $\alpha$.
The top pole mass has been taken as $m_t=172.6$ GeV, while the bottom
quark pole mass has been chosen to be $\mb=4.60$~GeV, which corresponds
to a $\MSbar$ mass $\overline{m}_b(\overline{m}_b)=4.26$~GeV. The strong
coupling constant has been normalized to $\alpha_s(M_Z)=0.118$. We have
used the top and bottom pole masses in the mass matrices of the stop
and sbottom states as implemented in HDECAY.

\begin{figure}
\begin{center}
\vspace*{-3.0cm}
\epsfig{file=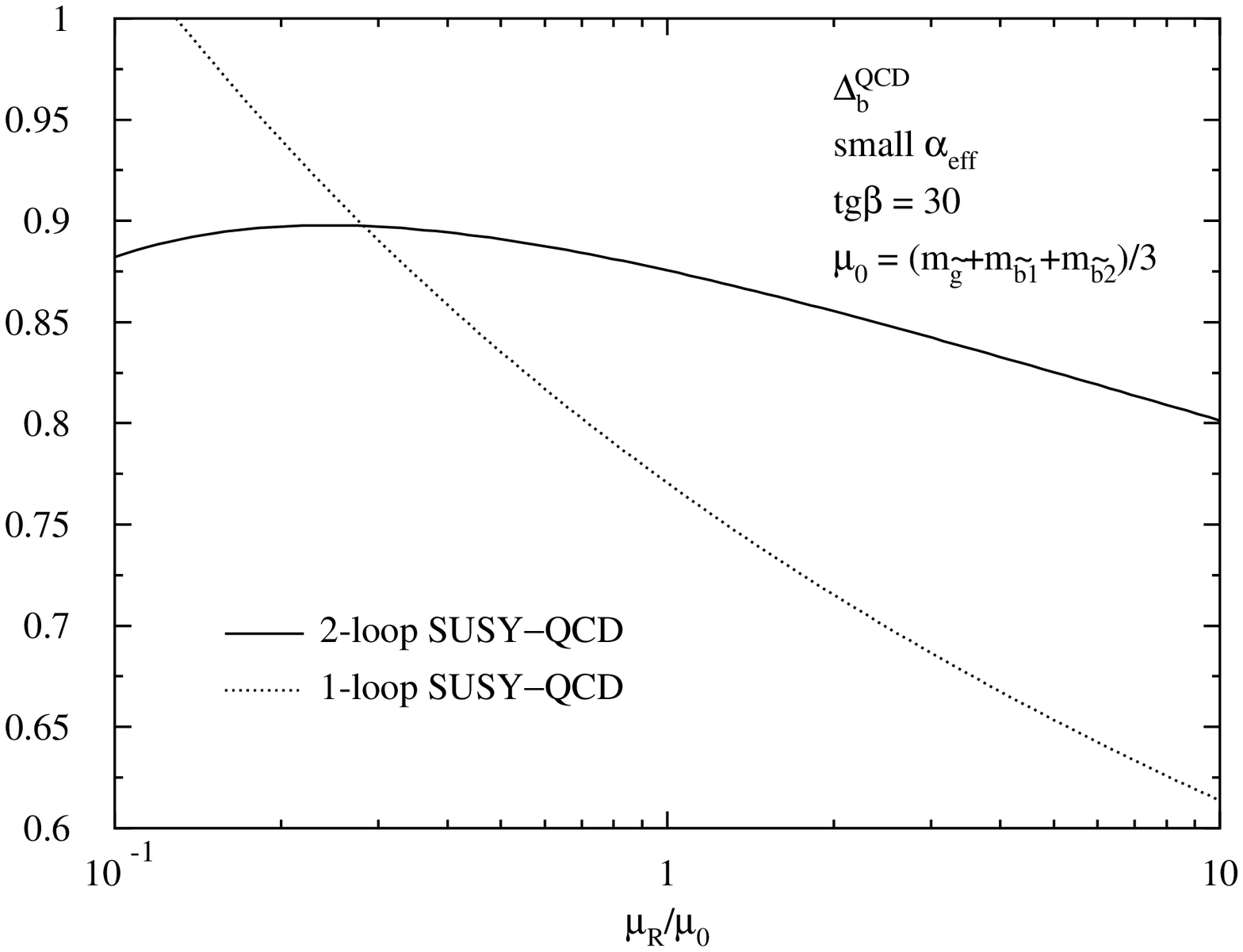,%
        bbllx=30pt,bblly=350pt,bburx=520pt,bbury=650pt,%
        scale=0.6}
\vspace*{2.5cm}

\epsfig{file=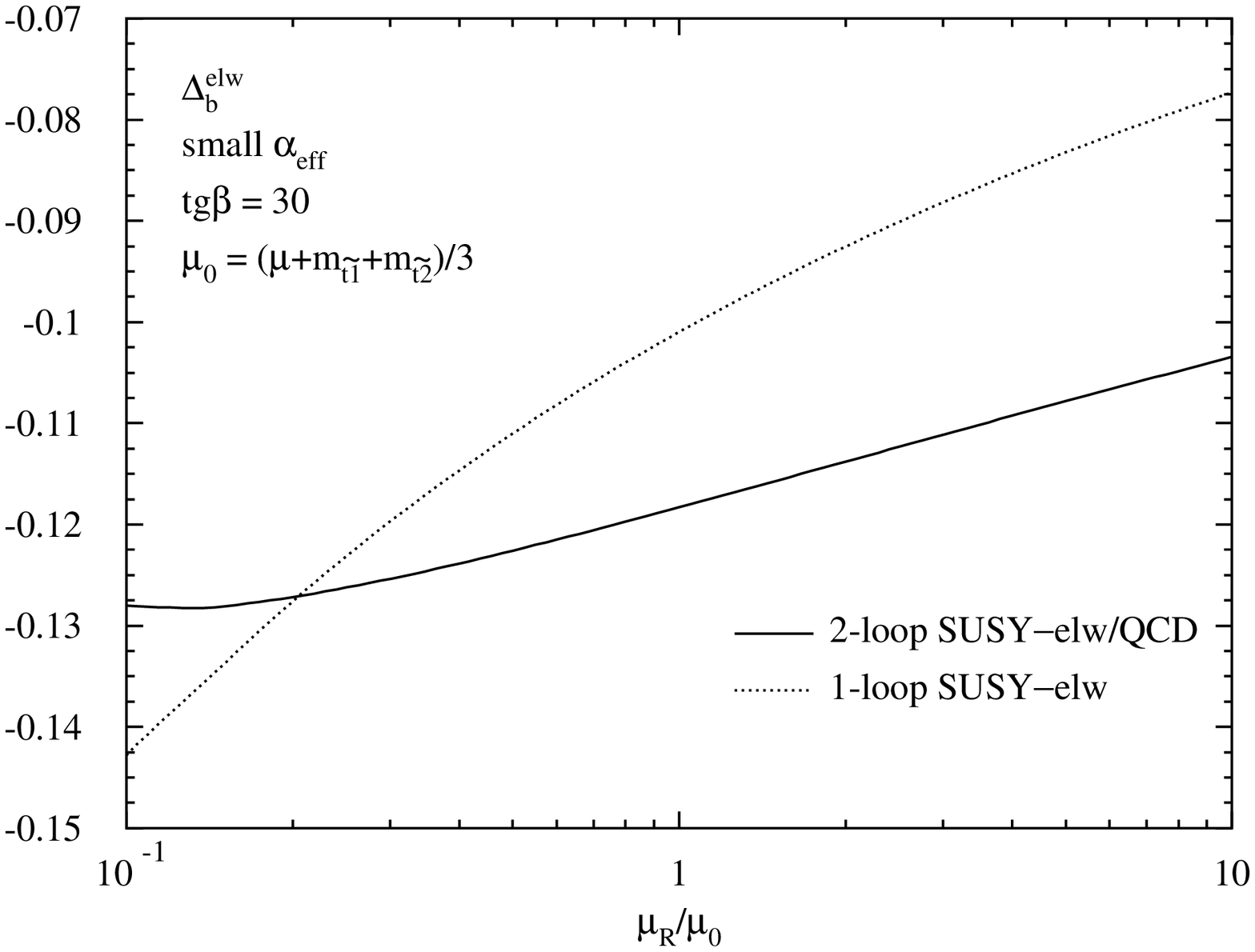,%
        bbllx=30pt,bblly=350pt,bburx=520pt,bbury=650pt,%
        scale=0.6}
\end{center}
\vspace*{2.0cm}
\caption{\it Scale dependence of the SUSY--QCD correction
$\Delta_b^{QCD}$ and the SUSY-electroweak correction $\Delta_b^{elw}$ at
one-loop and two-loop order in the small $\alpha_{eff}$ scenario.}
\label{fg:scale_smalla}
\end{figure}

\begin{figure}
\begin{center}
\vspace*{-3.0cm}
\epsfig{file=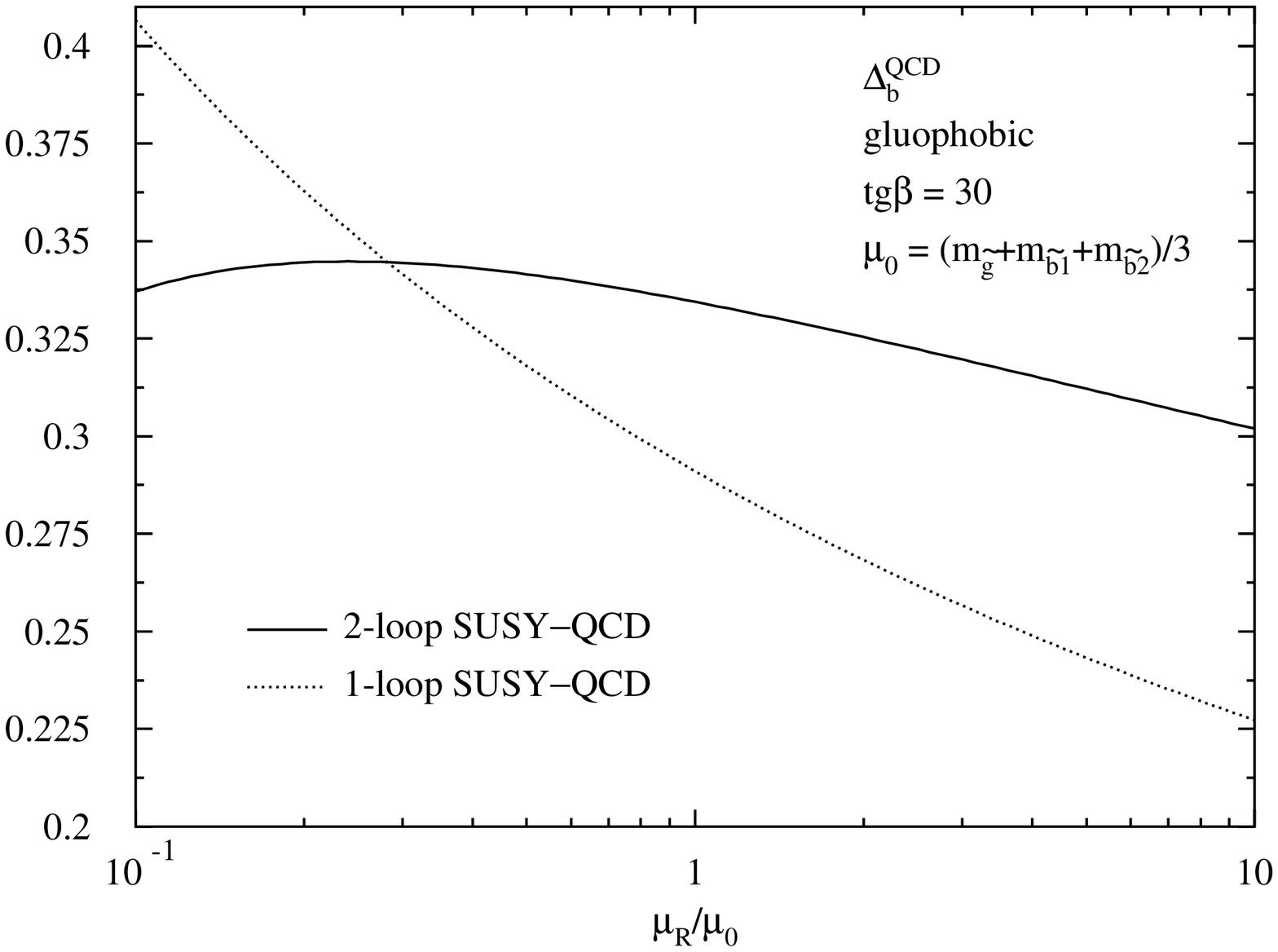,%
        bbllx=30pt,bblly=350pt,bburx=520pt,bbury=650pt,%
        scale=0.6}
\vspace*{2.5cm}

\epsfig{file=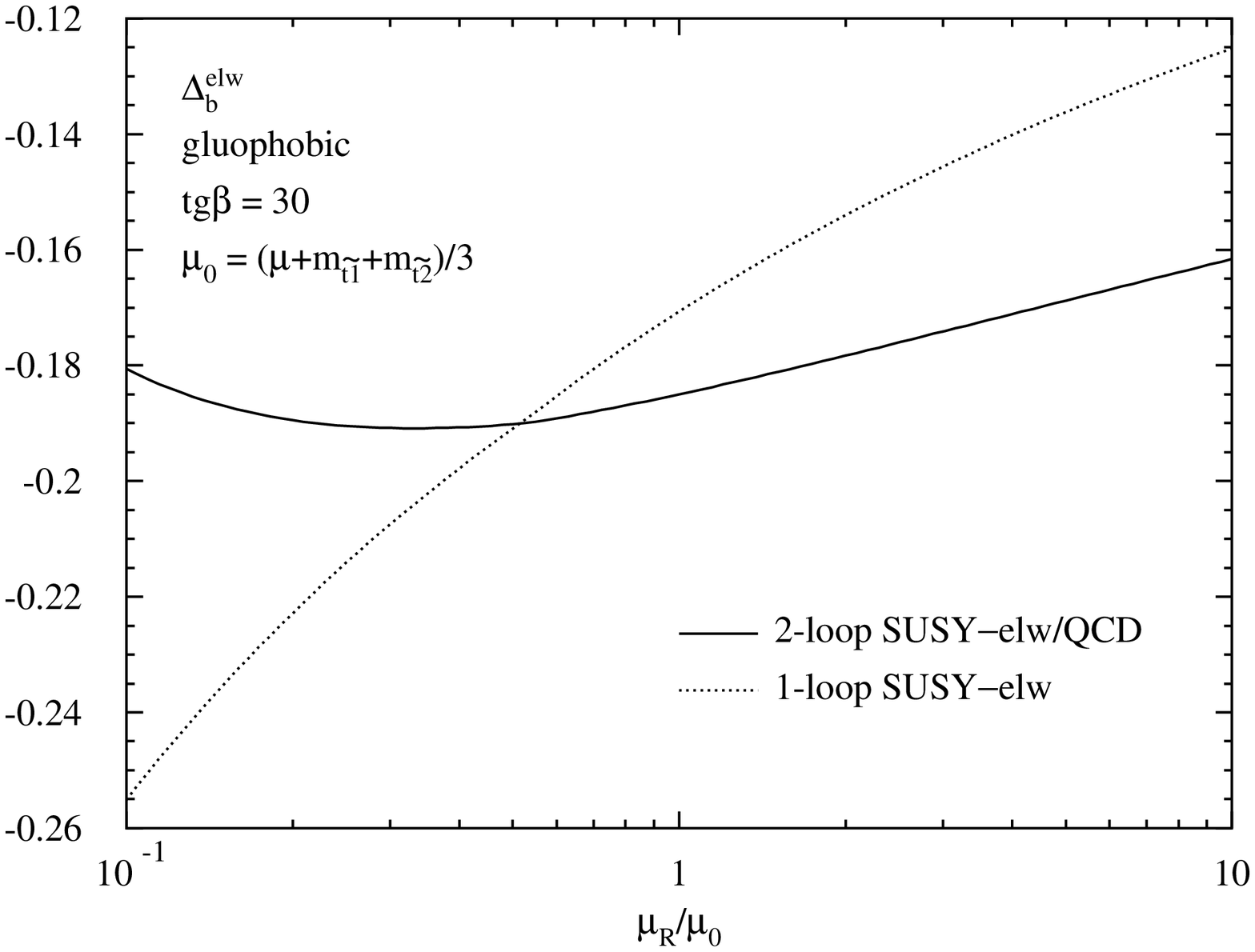,%
        bbllx=30pt,bblly=350pt,bburx=520pt,bbury=650pt,%
        scale=0.6}
\end{center}
\vspace*{2.0cm}
\caption{\it Scale dependence of the SUSY--QCD correction
$\Delta_b^{QCD}$ and the SUSY-electroweak correction $\Delta_b^{elw}$ at
one-loop and two-loop order in the gluophobic scenario.}
\label{fg:scale_gluoph}
\end{figure}

In the following we will not address the parametric uncertainties of the
corrections to the bottom Yukawa couplings and branching ratios, which
will be significantly reduced after the LHC and a linear $e^+e^-$
collider with energies in the range up to about 1 TeV \cite{ilc}.
However, we will concentrate mainly on the renormalization scale
dependence of our final results as a first estimate of the residual
theoretical uncertainties after including our novel NNLO corrections to
the bottom Yukawa couplings\footnote{The scheme and scale choices of the
stop, sbottom and gluino masses as well as the trilinear coupling $A_t$
modify the results by a few per cent at NLO already so that their impact
ranges below the per-cent level at NNLO. These effects are neglected in
our analysis, since the dominant uncertainties originate from the scale
choices of the strong coupling constant $\alpha_s$ and the top Yukawa
coupling $\lambda_t$.}. In
Fig.~\ref{fg:scale_smalla} the scale dependence of the $\Delta_b^{QCD}$
and $\Delta_b^{elw}$ terms is shown for the small $\alpha_{eff}$
scenario and in Fig.~\ref{fg:scale_gluoph} for the gluophobic scenario.
The $\Delta_b^{QCD}$ correction amounts to about 90\% in the small
$\alpha_{eff}$ scenario partially compensated by a negative
$\Delta_b^{elw}$ contribution of ${\cal O}(10\%)$. By comparing the
dashed one-loop bands with the full two-loop curves a significant
reduction of the residual scale dependence can be observed leading to a
final scale uncertainty in the per-cent range at NNLO. An analogous
reduction of the scale dependence can be observed in the gluophobic
scenario, where the $\Delta_b^{QCD}$ terms amount to about 35\% and are
significantly reduced by about 20\% due to the electroweak
$\Delta_b^{elw}$ contribution. In general, however, the sizes and signs
of these contributions strongly depend on the MSSM scenario. Moreover,
broad maxima/minima develop at about 1/3 of the central scales
determined by the average of the corresponding SUSY-masses,
i.e.~$\mu_0=(m_{\sbottom_1} + m_{\sbottom_2} + m_{\sgl})/3$ for
$\Delta_b^{QCD}$ and $\mu_0=(m_{\st_1} + m_{\st_2} + \mu)/3$ for
$\Delta_b^{elw}$ at NNLO in contrast to the monotonic scale dependence
of the one-loop expressions. For the central scale choices the two-loop
corrections to $\Delta_b^{QCD}$ amount up to ${\cal O}(10\%)$, while
they are in the per-cent range for $\Delta_b^{elw}$ \cite{noth}. Since
the corrections $\Delta_b$ enter the effective Yukawa couplings as in
Eq.~(\ref{eq:rescoup}) they have an immediate impact on all processes
induced by these bottom Yukawa couplings.

\begin{figure}
\begin{center}
\vspace*{-0.3cm}

\hspace*{-1.3cm}
\epsfig{file=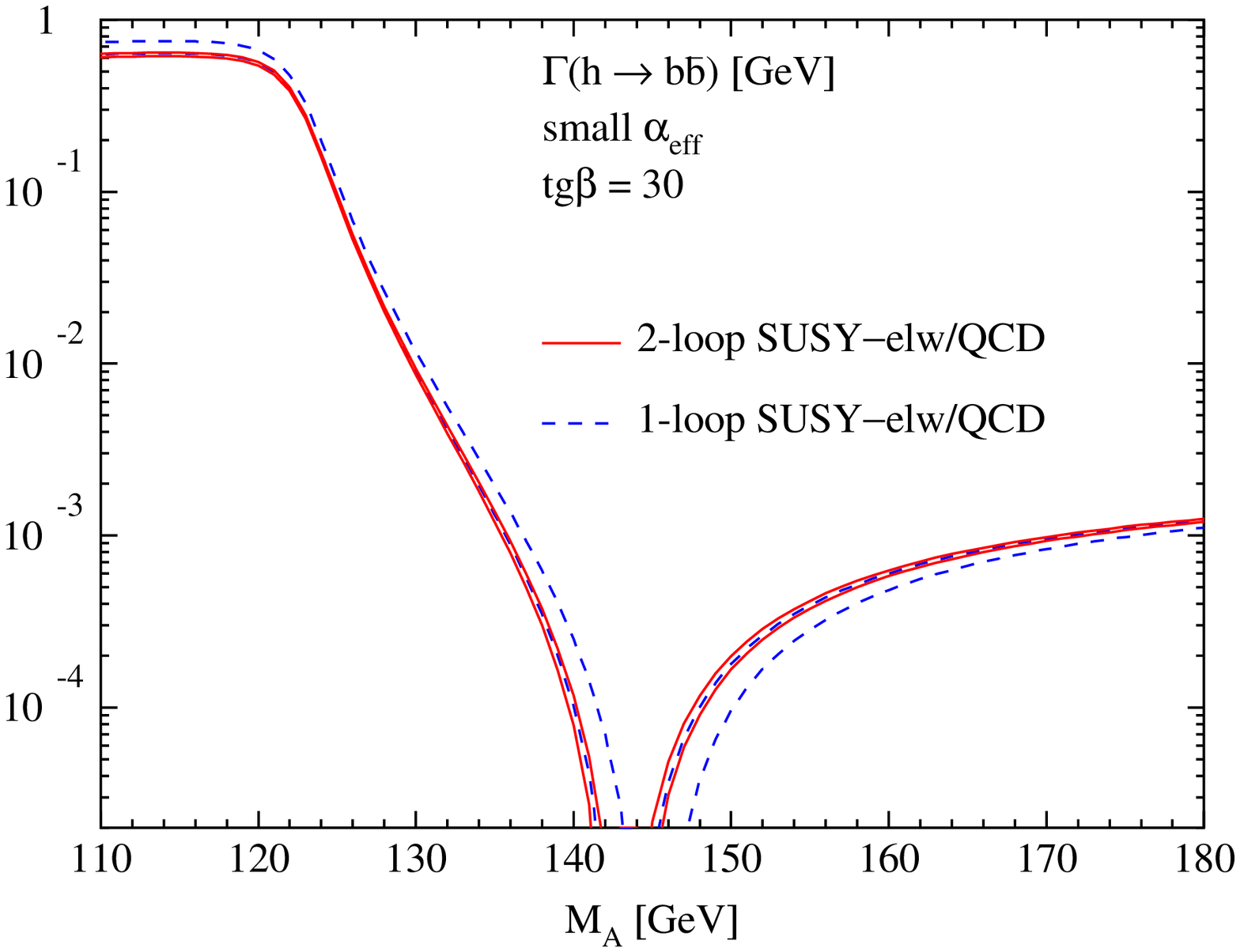,%
        bbllx=30pt,bblly=350pt,bburx=520pt,bbury=650pt,%
        scale=0.45}
\vspace*{1.7cm}

\hspace*{-1.3cm}
\epsfig{file=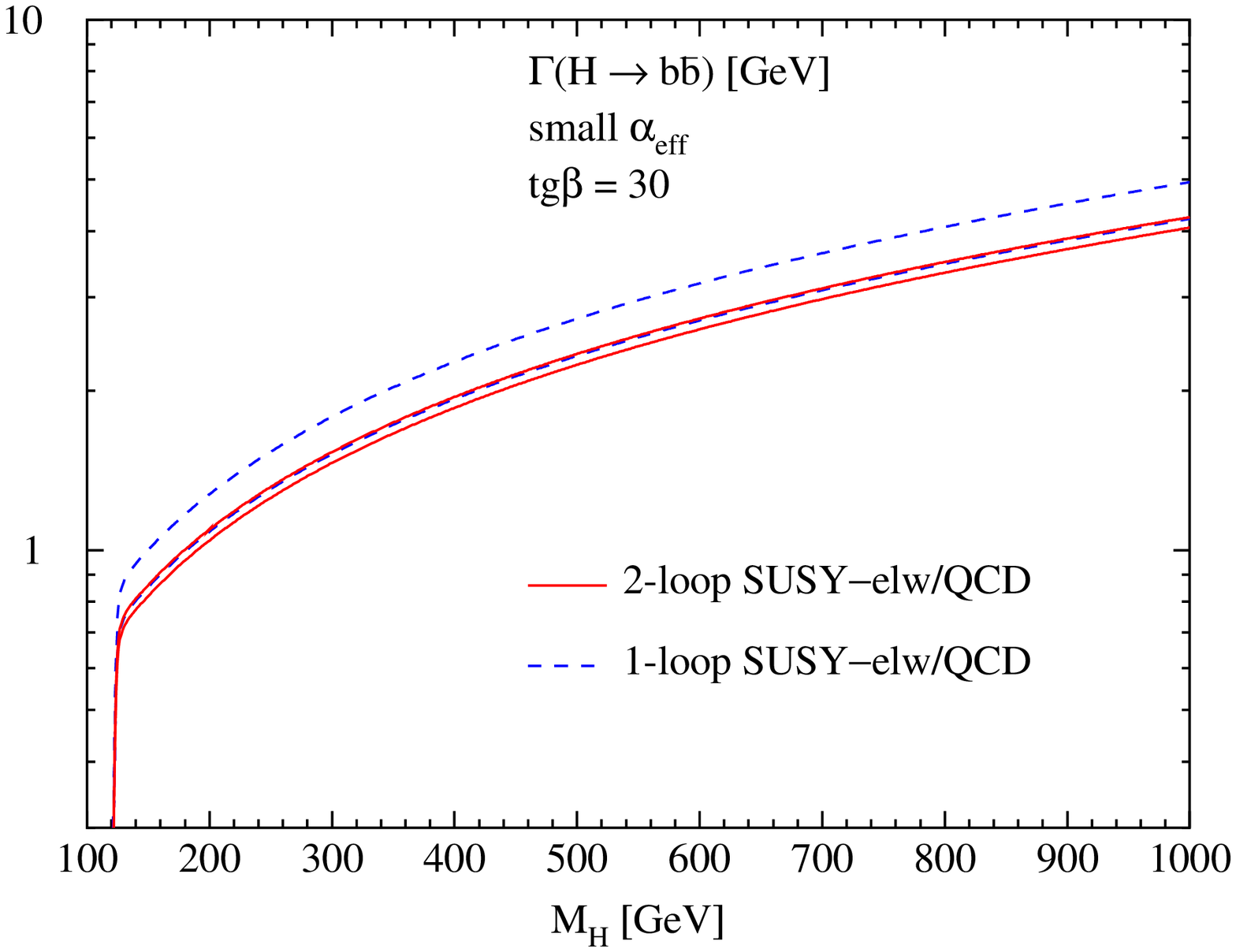,%
        bbllx=30pt,bblly=350pt,bburx=520pt,bbury=650pt,%
        scale=0.45}
\vspace*{1.7cm}

\hspace*{-1.3cm}
\epsfig{file=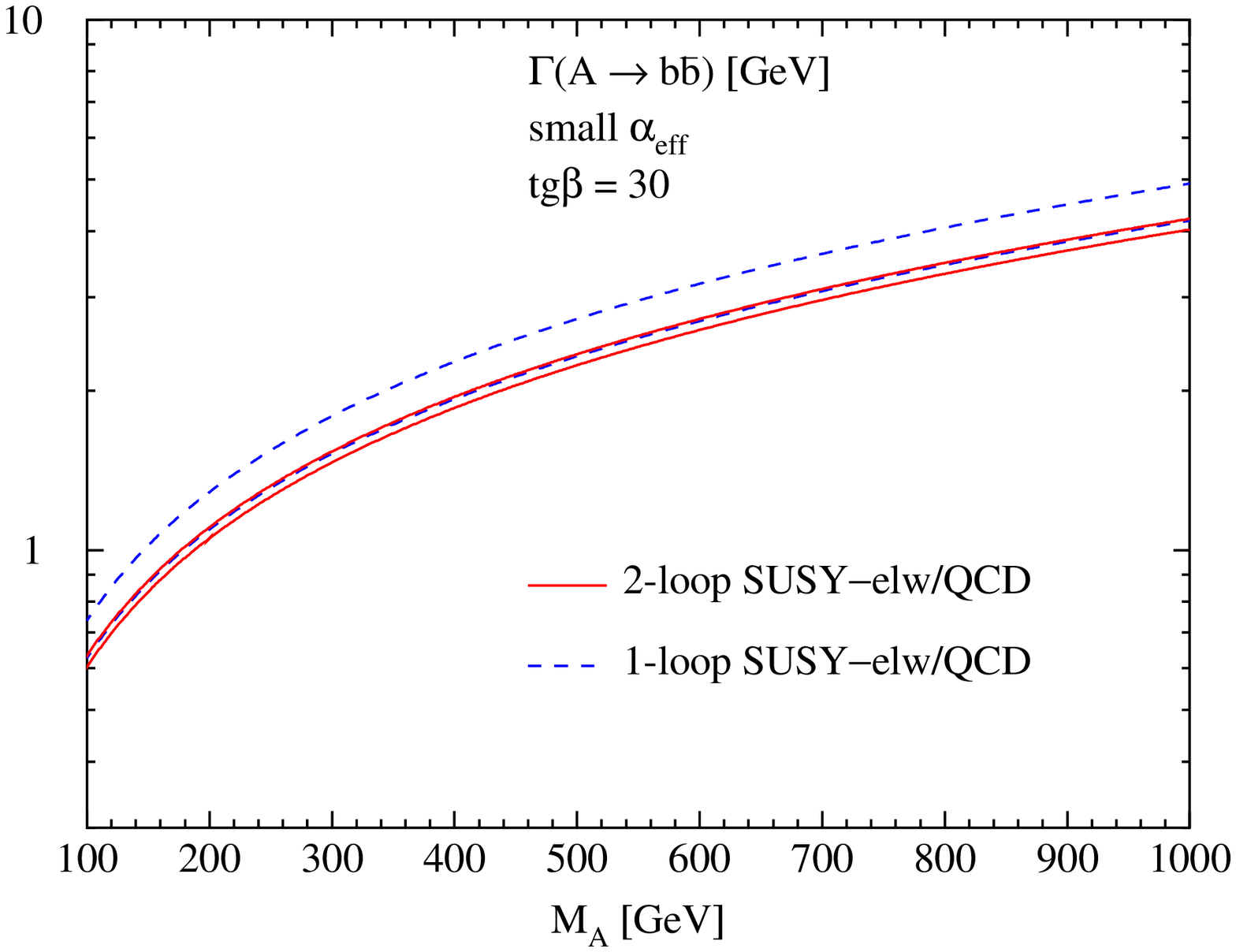,%
        bbllx=30pt,bblly=350pt,bburx=520pt,bbury=650pt,%
        scale=0.45}
\end{center}
\vspace*{1.2cm}
\caption{\it Partial decay widths of the light scalar $h$, the heavy
scalar $H$ and the pseudoscalar $A$ Higgs bosons to $b\bar b$ in the
small $\alpha_{eff}$ scenario. The dashed blue bands indicate the scale
dependence at one-loop order and the full red bands at two-loop order by
varying the renormalization scales of $\Delta_b^{QCD}$ and
$\Delta_b^{elw}$ independently between $1/2$ and $2$ times the central
scale given by the corresponding average of the SUSY-particle masses.}
\label{fg:Gamma_smalla}
\end{figure}

\begin{figure}
\begin{center}
\vspace*{-0.3cm}

\hspace*{-1.3cm}
\epsfig{file=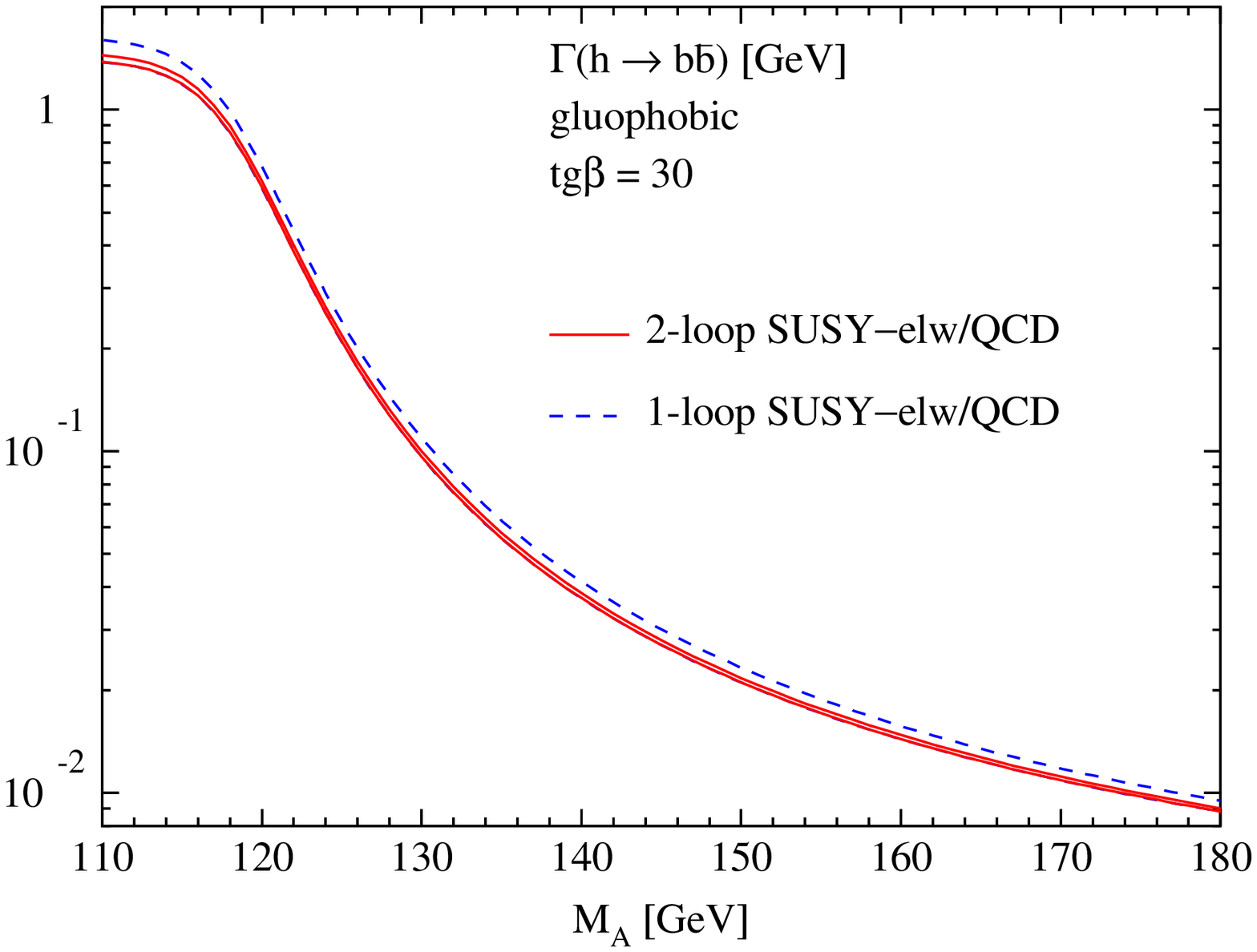,%
        bbllx=30pt,bblly=350pt,bburx=520pt,bbury=650pt,%
        scale=0.45}
\vspace*{1.7cm}

\hspace*{-1.3cm}
\epsfig{file=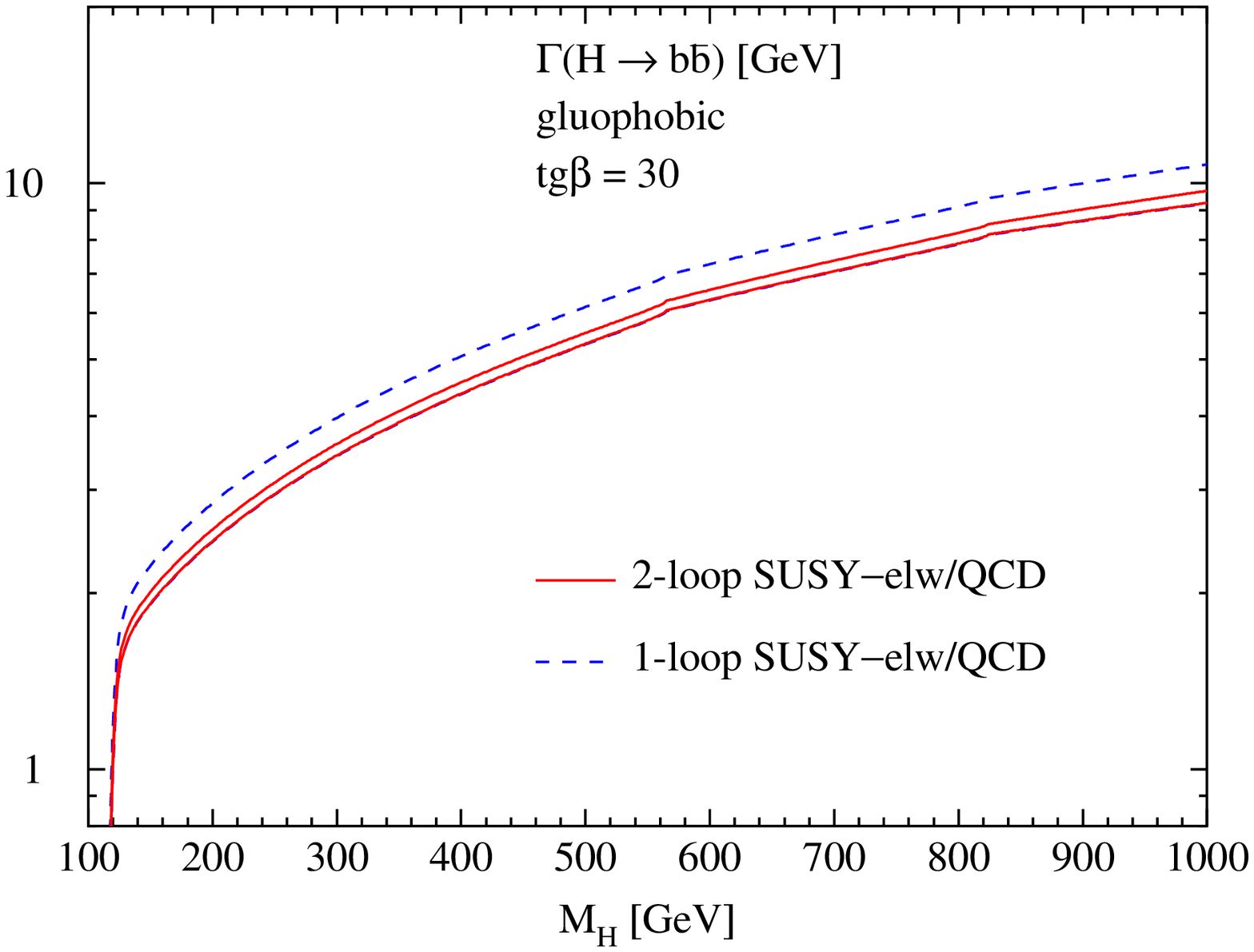,%
        bbllx=30pt,bblly=350pt,bburx=520pt,bbury=650pt,%
        scale=0.45}
\vspace*{1.7cm}

\hspace*{-1.3cm}
\epsfig{file=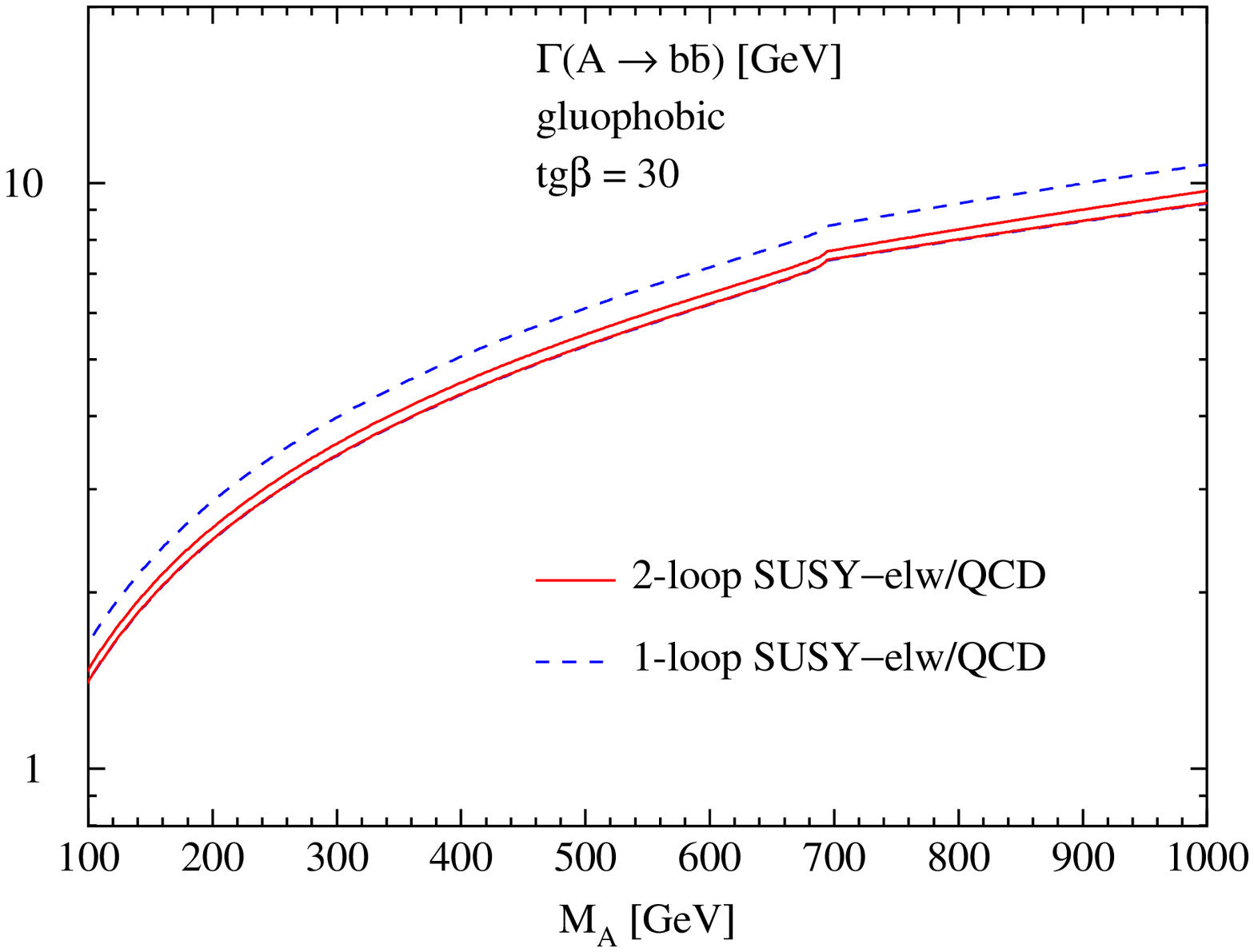,%
        bbllx=30pt,bblly=350pt,bburx=520pt,bbury=650pt,%
        scale=0.45}
\end{center}
\vspace*{1.2cm}
\caption{\it Partial decay widths of the light scalar $h$, the heavy
scalar $H$ and the pseudoscalar $A$ Higgs bosons to $b\bar b$ in the
gluophobic scenario. The dashed blue bands indicate the scale dependence
at one-loop order and the full red bands at two-loop order by varying
the renormalization scales of $\Delta_b^{QCD}$ and $\Delta_b^{elw}$
independently between $1/2$ and $2$ times the central scale given by the
corresponding average of the SUSY-particle masses.}
\label{fg:Gamma_gluoph}
\end{figure}

As a particular example we show the partial decay widths of the neutral
MSSM Higgs bosons into $b\bar b$ pairs in Figs.~\ref{fg:Gamma_smalla}
and \ref{fg:Gamma_gluoph} in the small $\alpha_{eff}$ and the gluophobic
scenario, respectively. The two-loop corrections to $\Delta_b$ lead to
negative corrections to the partial decay widths for the central scale
choices of ${\cal O}(10\%)$. The bands at one-loop order (dashed blue
curves) and two-loop order (full red curves) are defined by varying the
renormalization scales of $\Delta_b^{QCD}$ and $\Delta_b^{elw}$
independently between $1/2$ and $2$ times the corresponding central
scales $\mu_0$. A significant reduction of the dashed one-loop bands of
${\cal O}(10\%)$ to the full two-loop bands at the per-cent level can be
inferred from these results. The small kinks in the partial decay width
$\Gamma(H\to b\bar b)$ in the gluophobic scenario originate from the
$\sbottom_1 \bar\sbottom_1$ and the $\sbottom_2 \bar\sbottom_2$
thresholds in $\Delta_{SQCD}^{rem}$ of Eq.~(\ref{eq:gambb}). The small
kink in the partial decay width $\Gamma(A\to b\bar b)$ on the other hand
can be traced back to the $\sbottom_1 \bar\sbottom_2$ and $\sbottom_2
\bar\sbottom_1$ thresholds. The light scalar Higgs boson decay width
$\Gamma(h\to b\bar b)$ exhibits a strong dependence on the pseudoscalar
mass $M_A$ for $M_A\gtrsim 120$ GeV, since it slowly approaches the
SM-limit for large $M_A$. The drop of $\Gamma(h\to b\bar b)$ for
$M_A\sim 145$ GeV in Fig.~\ref{fg:Gamma_smalla} is due to the vanishing
of the effective mixing angle $\alpha$ in the MSSM Higgs sector.

\begin{figure}
\begin{center}
\vspace*{-0.3cm}

\hspace*{-1.3cm}
\epsfig{file=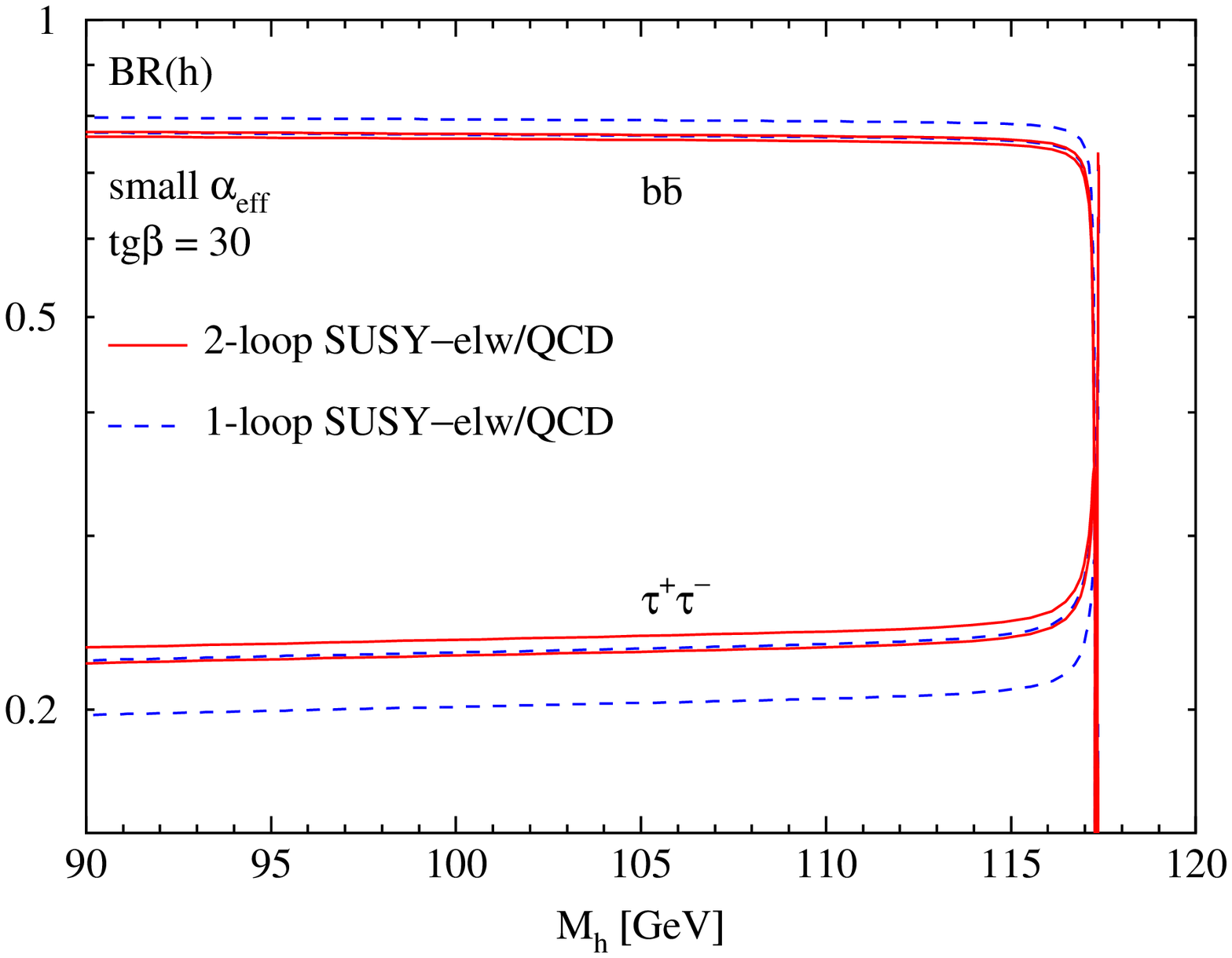,%
        bbllx=30pt,bblly=350pt,bburx=520pt,bbury=650pt,%
        scale=0.45}
\vspace*{1.7cm}

\hspace*{-1.3cm}
\epsfig{file=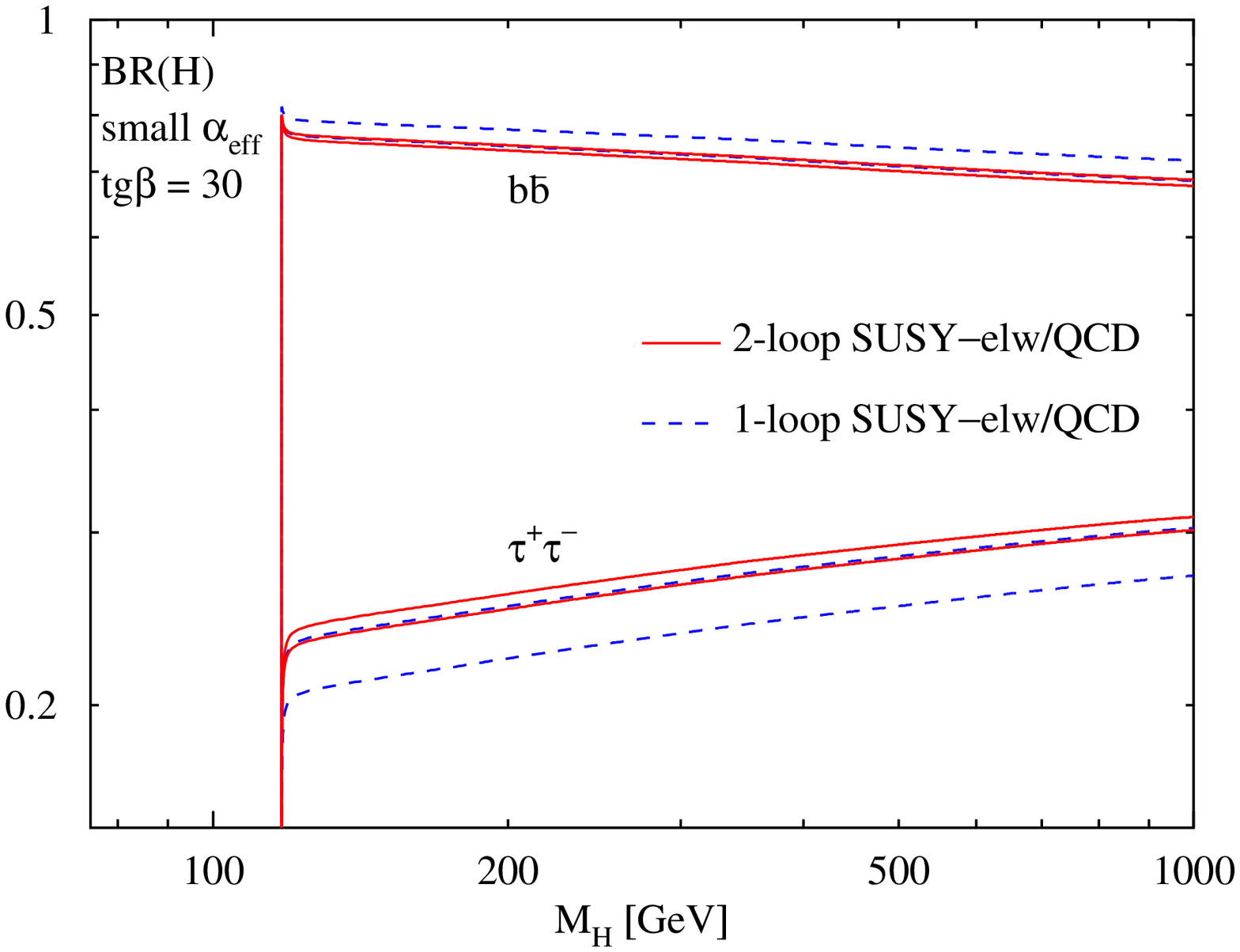,%
        bbllx=30pt,bblly=350pt,bburx=520pt,bbury=650pt,%
        scale=0.45}
\vspace*{1.7cm}

\hspace*{-1.3cm}
\epsfig{file=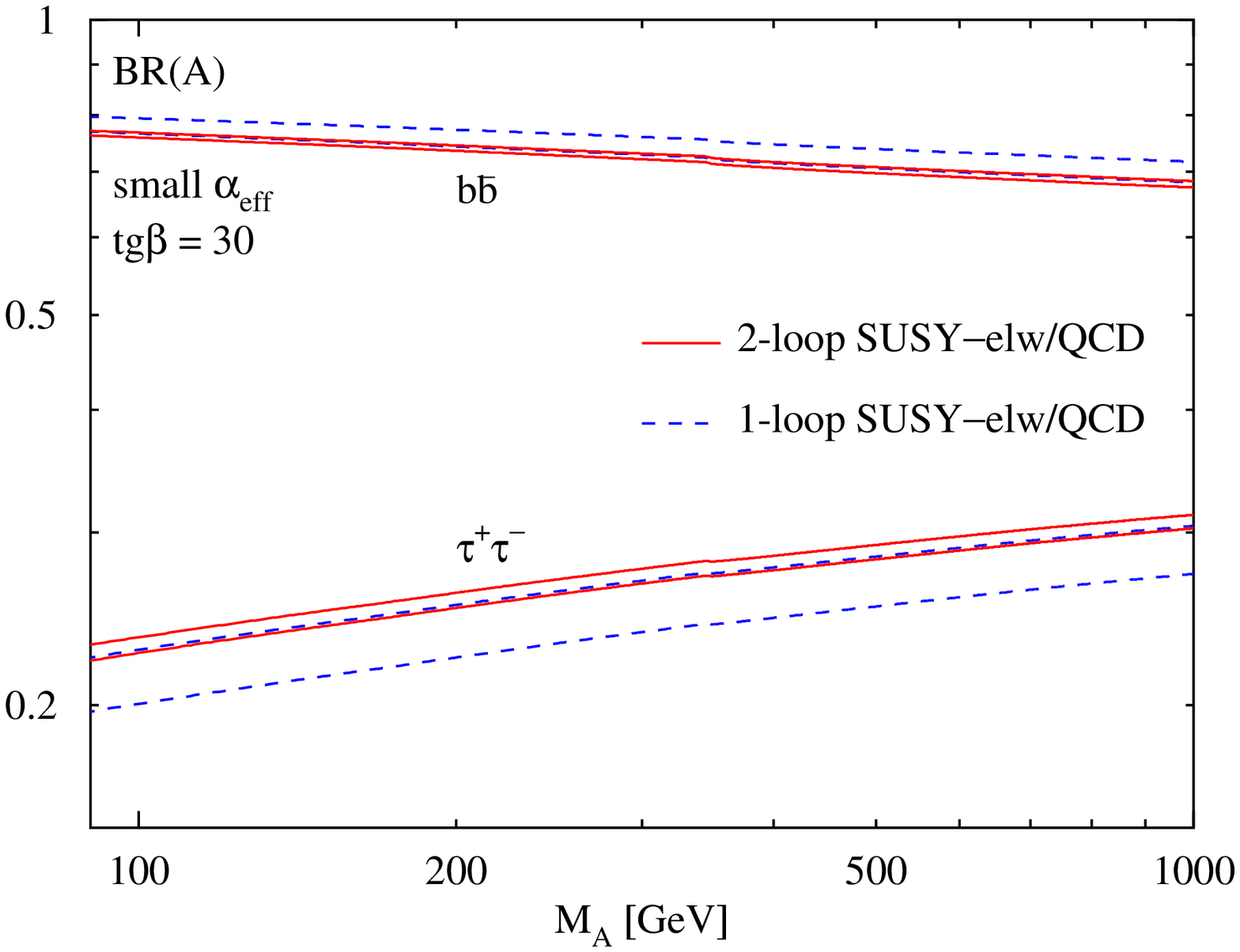,%
        bbllx=30pt,bblly=350pt,bburx=520pt,bbury=650pt,%
        scale=0.45}
\end{center}
\vspace*{1.2cm}
\caption{\it Branching ratios of the light scalar $h$, the heavy scalar
$H$ and the pseudoscalar $A$ Higgs bosons to $b\bar b$ and
$\tau^+\tau^-$ in the small $\alpha_{eff}$ scenario. The dashed blue
bands indicate the scale dependence at one-loop order and the full red
bands at two-loop order by varying the renormalization scales of
$\Delta_b^{QCD}$ and $\Delta_b^{elw}$ independently between $1/2$ and
$2$ times the central scale given by the corresponding average of the
SUSY-particle masses.}
\label{fg:BR_smalla}
\end{figure}

\begin{figure}
\begin{center}
\vspace*{-0.3cm}

\hspace*{-1.3cm}
\epsfig{file=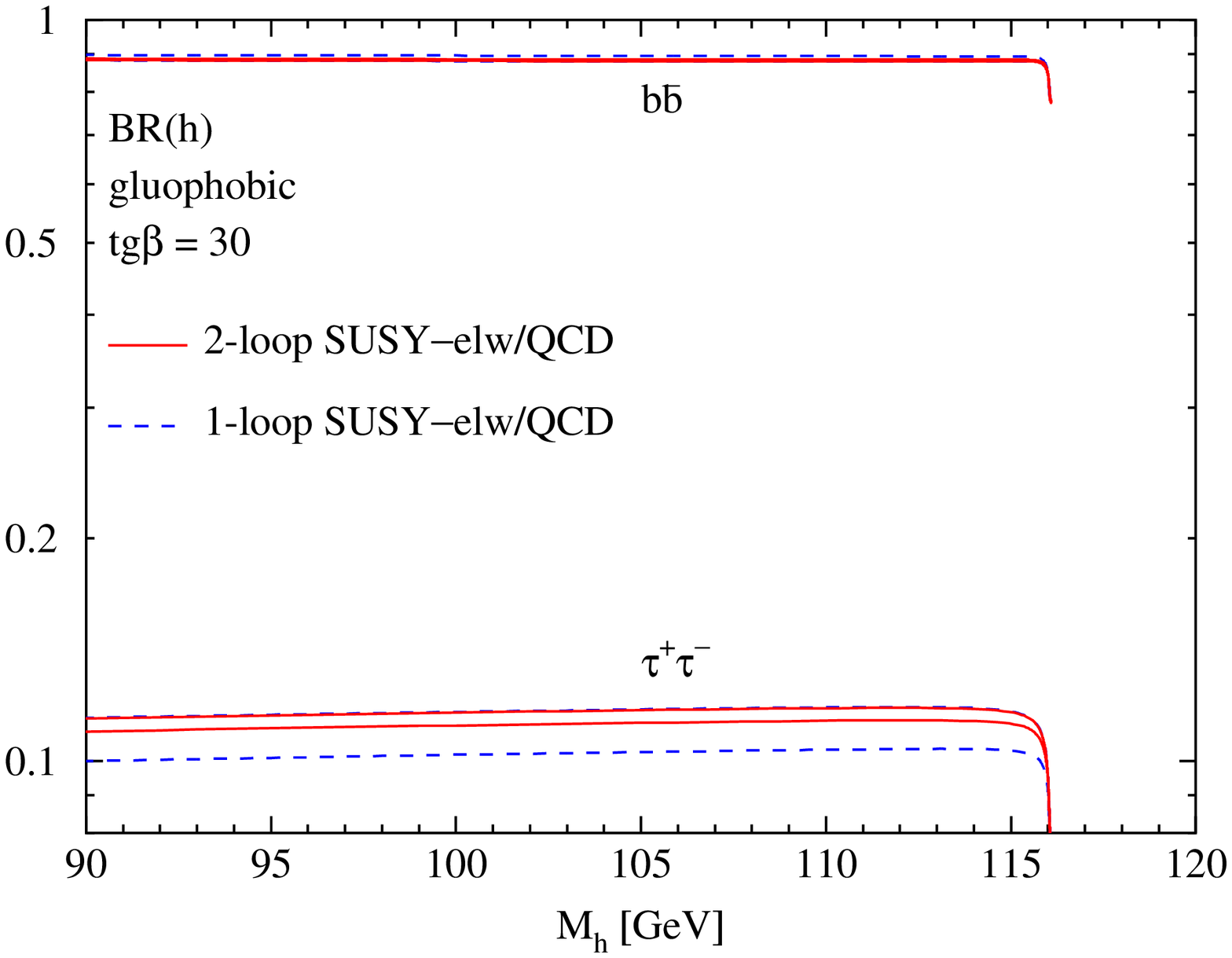,%
        bbllx=30pt,bblly=350pt,bburx=520pt,bbury=650pt,%
        scale=0.45}
\vspace*{1.7cm}

\hspace*{-1.3cm}
\epsfig{file=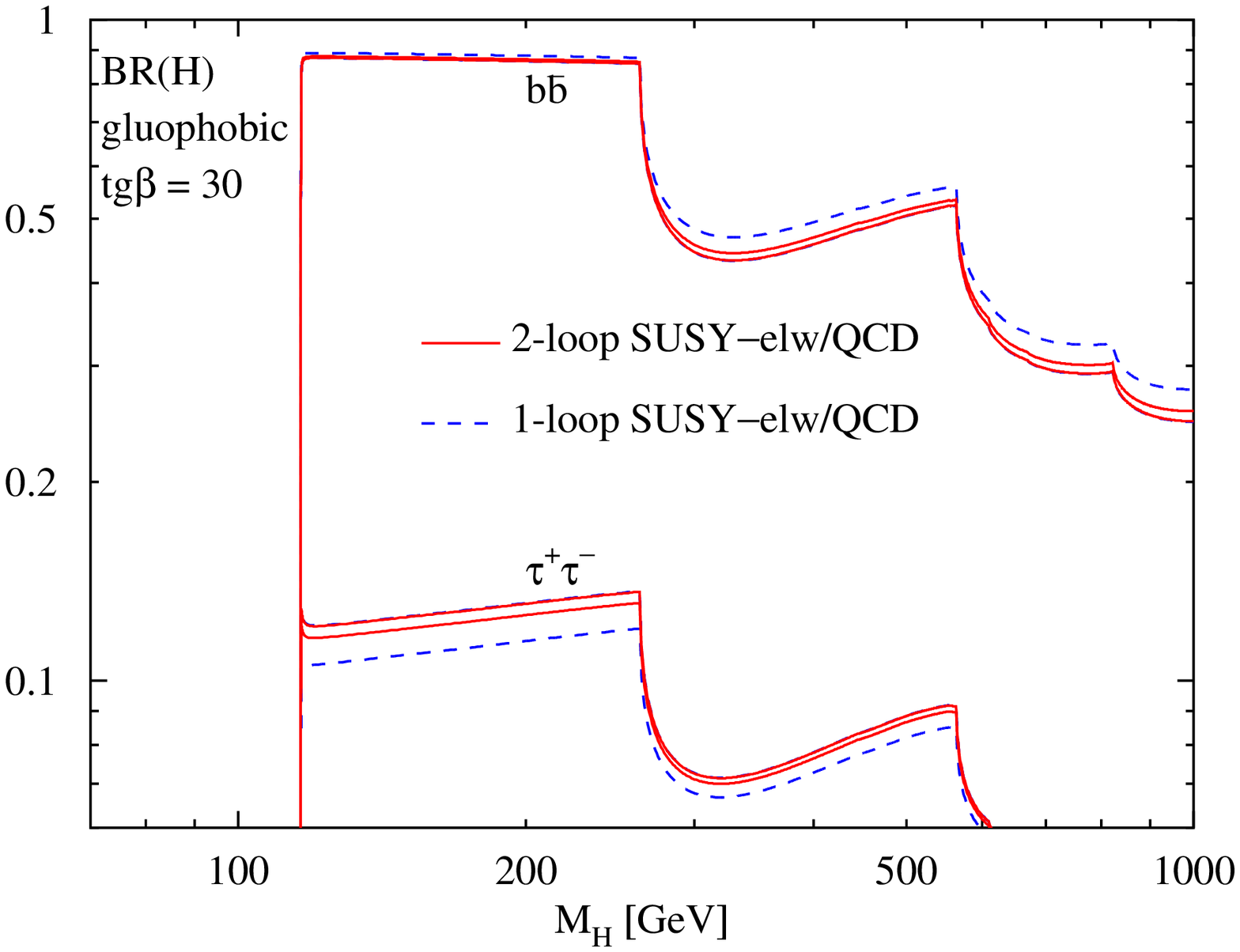,%
        bbllx=30pt,bblly=350pt,bburx=520pt,bbury=650pt,%
        scale=0.45}
\vspace*{1.7cm}

\hspace*{-1.3cm}
\epsfig{file=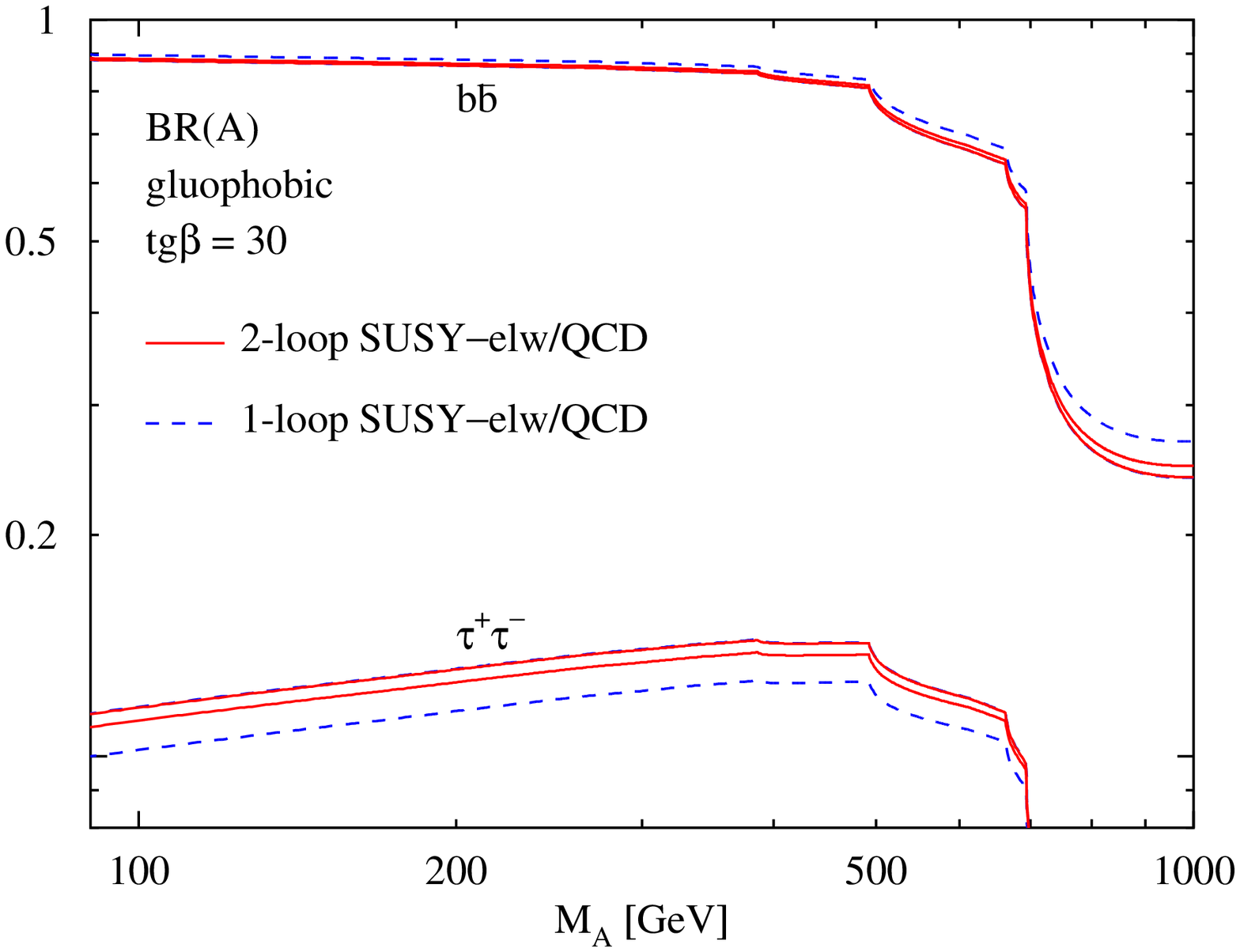,%
        bbllx=30pt,bblly=350pt,bburx=520pt,bbury=650pt,%
        scale=0.45}
\end{center}
\vspace*{1.2cm}
\caption{\it Branching ratios of the light scalar $h$, the heavy scalar
$H$ and the pseudoscalar $A$ Higgs bosons to $b\bar b$ and
$\tau^+\tau^-$ in the gluophobic scenario. The dashed blue bands
indicate the scale dependence at one-loop order and the full red bands
at two-loop order by varying the renormalization scales of
$\Delta_b^{QCD}$ and $\Delta_b^{elw}$ independently between $1/2$ and
$2$ times the central scale given by the corresponding average of the
SUSY-particle masses.}
\label{fg:BR_gluoph}
\end{figure}

The scale uncertainties in the partial decay widths translate into
theoretical uncertainties in the corresponding branching ratios of the
neutral MSSM Higgs bosons. These errors cancel to a large extent in the
dominant branching ratio $BR(\Phi\ra\bbbar)$, but induce sizable
uncertainties of the non-leading branching ratios at one-loop order. The
branching ratios for the two dominant decay modes into $b\bar b$ and
$\tau^+\tau^-$ pairs are depicted in Figs.~\ref{fg:BR_smalla} and
\ref{fg:BR_gluoph} for the small $\alpha_{eff}$ and the gluophobic
scenario, respectively.  The uncertainties of the branching ratios
reduce from $\Order{10\%}$ at one-loop order to the per-cent level at
two-loop order. The per-cent accuracy now matches the expected
experimental accuracies at a future linear $e^+e^-$ collider. The
thresholds visible in the branching ratios of the heavy scalar Higgs
boson in Fig.~\ref{fg:BR_gluoph} are due to the kinematical opening of
the $\st_1\bar \st_1, \sbottom_1 \bar\sbottom_1$ and $\sbottom_2
\bar\sbottom_2$ decay modes in consecutive order with rising Higgs mass.
The thresholds of the pseudoscalar branching ratios are due to the decay
modes into $\tilde \chi_1^0\tilde \chi_2^0, \tilde \chi_1^+\tilde
\chi_1^-, \st_1 \st_2$ and $\sbottom_1 \sbottom_2$.

Since we have determined the effective resummed bottom Yukawa couplings
at two-loop order the results will also affect all other processes which
are significantly induced by bottom Yukawa couplings, as e.g. MSSM Higgs
radiation off bottom quarks at $e^+e^-$ colliders \cite{ee2hbb,eebbh} and
hadron colliders \cite{ppbbh}.  The two-loop corrections can easily be
included in the corresponding numerical programs. However, these
analyses are beyond the scope of this work.

\section{Conclusions}
In this work we have determined the NNLO corrections to the effective
bottom quark Yukawa couplings within the MSSM for large values of
$\tgb$, since the theoretical uncertainties are sizable at NLO in these
regimes. The leading parts of the SUSY-QCD and top-induced
SUSY-electroweak corrections originate from factorizable contributions
due to virtual squark, gluino and Higgsino exchange, which can be
absorbed in effective Yukawa couplings in a universal way. We have
calculated the two-loop SUSY-QCD corrections to these effective bottom
Yukawa couplings.

In summary, the significant scale dependence of $\Order{10\%}$ of the
NLO predictions for processes involving the bottom quark Yukawa
couplings of supersymmetric Higgs bosons require the inclusion of NNLO
corrections. For the corrected Yukawa couplings, we find a reduction of
the scale dependence to the per-cent level at the NNLO level.  The
improved NNLO predictions for the bottom Yukawa couplings can thus be
taken as a base for experimental analyses at the Tevatron and the LHC as
well as the ILC. \\

\noindent
{\bf Note added} \\
During the publication procedure of our work another paper appeared
\cite{reisser} where the bottom Yukawa couplings have been calculated at
two-loop order within SUSY--QCD. After adjusting the different
renormalization schemes and approximations we found mutual agreement
between the two calculations. \\

\noindent
{\bf Acknowledgments} \\
The authors are pleased to thank J.~Guasch for many useful discussions
and carefully reading the manuscript. We are indebted to
M.M.~M\"uhlleitner, H.~Rzehak and P.M.~Zerwas for comments on the
manuscript. We would like to thank L.~Mihaila for a detailed comparison
of the results of Ref.~\cite{reisser} with ours. This work is
supported in part by the the Swiss National Science Foundation and the
European Community's Marie-Curie Research Training Network HEPTOOLS
under contract MRTN-CT-2006-035505.



\end{document}